\documentclass{amsart}

\pdfoutput=1 
\usepackage{graphicx}
\usepackage{color,soul}
\usepackage[table]{xcolor}
\graphicspath{ {Images/} }
\usepackage{caption}
\usepackage{subcaption}
\usepackage{comment}

\theoremstyle{definition}

\theoremstyle{remark}

\definecolor{col1}{rgb}{0.3,0.3,0.3}
\definecolor{col2}{rgb}{0.5,0.5,0.5}
\newcommand{\au}[1]{
    \ifnum#1<50
        \cellcolor{col1}
        \color{white}
    \else
        \ifnum#1<90
            \cellcolor{col2}
        \fi
    \fi
    #1\%}

\numberwithin{equation}{section}

\begin{document}

\title[Extending Hypoth. Testing with Persistence Hom.]{Extending
Hypothesis Testing with Persistence Homology to Three or More Groups.}

\author[C. Cericola]{Christopher Cericola}
\address{Christopher Cericola \\Department of Mathematics \\Louisiana State University\\ Baton Rouge, LA 70803}
\email{cceric1@lsu.edu}

\author[I. Johnson]{Inga Johnson}
\address{Inga Johnson \\ Department of Mathematics\\ Willamette University\\ Salem, OR 97301}
\email{ijohnson@willamette.edu}

\author[J. Kiers]{Joshua Kiers}
\address{Joshua Kiers\\
Department of Mathematics\\ University of North Carolina at Chapel Hill\\ Chapel Hill, NC 27599}
\email{jokiers@live.unc.edu}

\author[M. Krock]{Mitchell Krock}
\address{Mitchell Krock \\Department of Applied Mathematics\\ University of Colorado\\ Boulder, CO 80309}
\email{Mitchell.Krock@Colorado.EDU}

\author[J. Purdy]{Jordan Purdy}
\address{Jordan Purdy \\Department of Mathematics \\Willamette University \\Salem, OR 97301}
\email{jpurdy@willamette.edu}

\author[J. Torrence]{Johanna Torrence}
\address{Johanna Torrence \\Department of Computer Science \\University of Chicago \\Chicago, IL 60637}
\email{jtorrence@uchicago.edu}

\thanks{This work was supported by an NSF DMS grant, \#1157105.}



\begin{abstract}
We extend the work of Robinson and Turner to use 
hypothesis testing with persistence homology to test for measurable differences in shape between point clouds from three or more groups.  Using samples of point clouds from three distinct groups, we conduct a large-scale simulation study to validate our proposed extension.  We consider various combinations of groups, samples sizes and measurement errors in the simulation study, providing for each combination the percentage of $p$-values below an alpha-level of 0.05.   Additionally, we apply our method to a Cardiotocography data set and find statistically significant evidence of measurable differences in shape between  normal, suspect and pathologic health status groups. 


\end{abstract}

\maketitle





\section{Introduction}
Consider a data set where each data point is a vector of $m$ quantitative variables and one categorical variable with $s$ levels. Ideally, several of the quantitative variables are real-valued.  According to the categorical variable, we will view the data set as $s$ not necessarily distinct collections of points in $\mathbb{R}^m$, referred to as point clouds. Additionally, assuming the data were randomly sampled, we will view each of these $s$ point clouds as a representative subset of their respective space.  Of interest is whether or not the spaces corresponding to these $s$ point clouds have measurably different shapes?  But what does shape even mean if $m$ is large? 

Topology, in particular algebraic topology, is an area of mathematics that can be used to  qualitatively measure the shape of a point cloud. For a given point cloud, we construct an infinite family of simplicial complexes that vary according to a real-valued distance parameter. Each complex in the family is an object that inherits a shape from the point cloud and the topological tool known as homology can be used to detect this shape.  Since any single complex within the infinite family corresponds to a choice of parameter value, we might ask which parameter value, if any, ``best" captures the shape of the point cloud?  Persistence homology is a study of the homological features that persist over long intervals of the distance parameter, thus sidestepping the search for a best choice  parameter value.  Hence, persistence homology can be used to determine if the $s$ point clouds in the data set have different shape.

While persistence homology allows comparisons of shape across the $s$ sampled point clouds, can any resulting sample differences then be generalized to the corresponding spaces at large?  The answer is yes, but as random sampling unavoidably introduces variability, a method is needed which can distinguish ``true" differences in shape between the spaces from ``artificial" differences between the sampled point clouds.  Statistical hypothesis testing is an inferential method often implemented to assess whether or not randomly sampled data provide sufficient evidence of a difference, with respect to some characteristic, between two or more populations, which we have been and will continue to loosely refer to as spaces.  In the 2013 results of K. Turner and A. Robinson, such an assessment is conducted on $s=2$ spaces using a specific type of hypothesis testing procedure known as a permutation test, where the characteristic of interest is shape~\cite{RobinsonTurner}.  In this procedure, the randomly sampled data are numerous point clouds from both spaces and the shape of a point cloud is measured via persistence homology.  In this paper we extend this procedure to three or more spaces, $s \geq 3$.  

The remainder of the paper is organized as follows.  In Section~\ref{hom} we provide definitions and examples of the Vietoris-Rips complex of a point cloud, homology groups, persistence homology and persistence diagrams.  In Section~\ref{HTandTDA2spaces} we describe the permutation test of Robinson and Turner.  In Section~\ref{HTandTDA3spaces} we propose an extension of the permutation test for three or more groups.  In Section~\ref{HTandTDA3spacesSimStudy} we present the results of a large-scale simulation study, incorporating various measurement errors and sample sizes, that validate our proposed extension.  Finally, in Section~\ref{CardioApp} we apply our extension to a Cardiotocography data set and find significant evidence of differences in shape, as measured by persistence homology, between the spaces corresponding to normal, suspect and pathologic health groups.~\footnote{
Throughout this paper we use the language difference in shape to mean shape as measured by persistence homology in a specified dimension.}

\section{Persistence Homology}\label{hom} 
Before defining the persistence homology of a point cloud, we associate to the point cloud a nested family of abstract simplicial complexes. A thorough explanation of simplicial complexes and abstract simplicial complexes is available in many sources~\cite{edelsbrunner, Munkres}.  Here we motivate the definition of an abstract simplicial complex with  a brief geometric introduction to simplicial complexes, followed by the definition of the Vietoris-Rips complex which is the abstract simplicial complex used herein. 

Geometrically, a 0-simplex is a point, a 1-simplex is a line segment, a 2-simplex is a triangular subset of a plane, a 3-simplex is a solid tetrahedron, and an $n$-simplex is the $n$-dimensional analogue of these convex sets. Observe that the boundary of an $n$-simplex, $\sigma$, is a collection of $(n-1)$-simplices; these boundary simplices are called faces of $\sigma$.  A {\it simplicial complex} is a collection of simplices in $\mathbb{R}^d$ that satisfy certain subset and intersection properties specifying how simplices can be put together to create a larger structure. More precisely, a simplicial complex is a finite collection of simplices, $K$, such that (1) if $\sigma \in K$ and $\rho$ is a face of $\sigma$ then $\rho \in K$, and (2) given any two simplices $\sigma_1, \sigma_2 \in K$ then $\sigma_1 \cap \sigma_2$ is either the empty set or a face of both $\sigma_1$ and $\sigma_2$.  More generally, and without relying on geometry, an {\it abstract simplicial complex} is a finite collection of sets, $A$, such that if $\alpha \in A$ and $\beta \subseteq \alpha$, then $\beta \in A$.  It is well known that a finite abstract simplicial complex can be geometrically realized as a simplicial complex in $\mathbb{R}^N$ for $N$ sufficiently large.

\subsection{The Vietoris-Rips Complex} 
The {\it Vietoris-Rips complex}, denoted $VR(D,r)$, is an abstract simplicial complex associated to a point cloud $D$ for a fixed radius value $r>0$. The elements of $D$ form the 0-simplices or vertex set of $VR(D,r)$. A simplex of $VR(D,r)$ is a finite subset $\alpha$ of $D$ such that the diameter of $\alpha$ is less than $r$.  A simplex $\alpha\subseteq D$ with $k$-elements is called a  $(k-1)$-simplex of $D$. Thus, a 1-simplex corresponds to a two element set (viewed geometrically as the endpoints of a line segment), a 2-simplex corresponds to a three element set (viewed as the vertices of a triangle), and so on.  Observe that if $\alpha$ is a $k$-simplex, then every subset of $\alpha$ is a simplex of $D$ as the diameter of a subset  of $\alpha$ can be no larger than the diameter of $\alpha$. Hence the Vietoris-Rips complex satisfies the definition of an abstract simplicial complex.

 \begin{figure}[h]\caption{Five data points in the plane.}\label{data5}
\centering
\includegraphics[width=.53\textwidth]{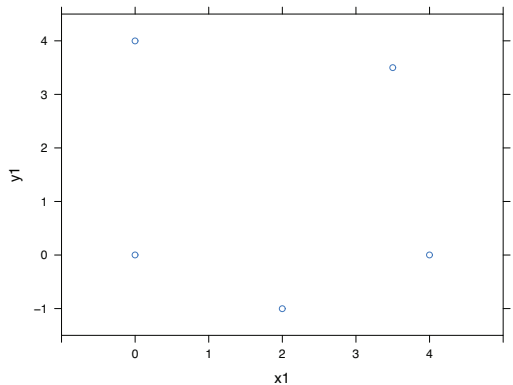}
\end{figure}

\begin{figure}[h]\caption{Representations of the abstract simplicial complexes $VR(D,4)$  and $VR(D,4.9)$ for a five point data set, $D$.}\label{VRexamples}
\centering
\includegraphics[scale=0.4]{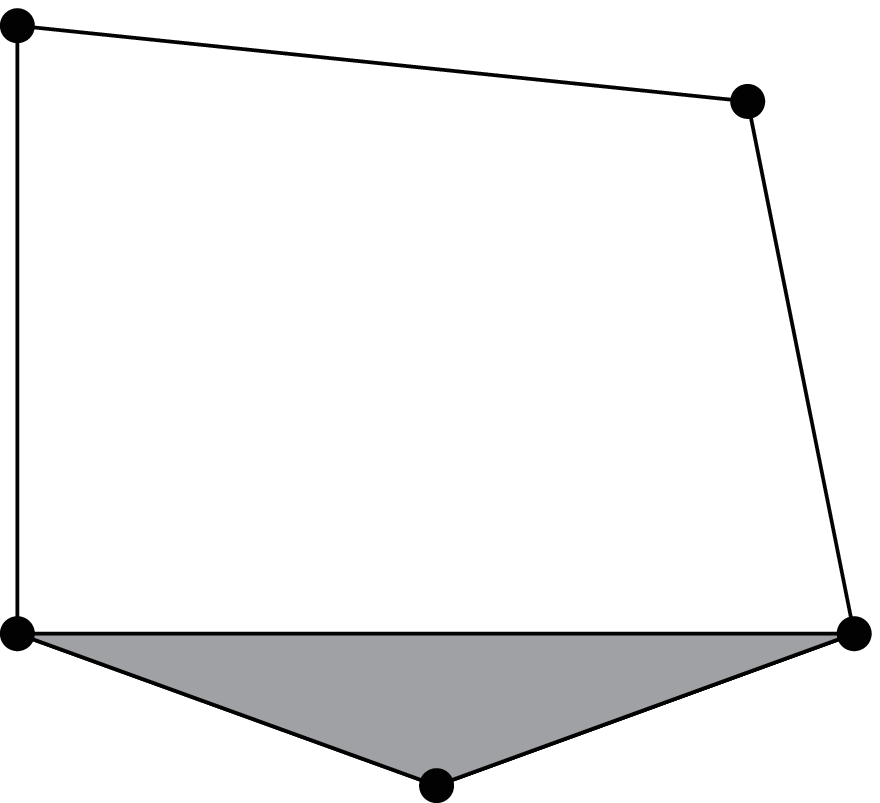}
\hspace{0.2in}
\includegraphics[scale=0.4]{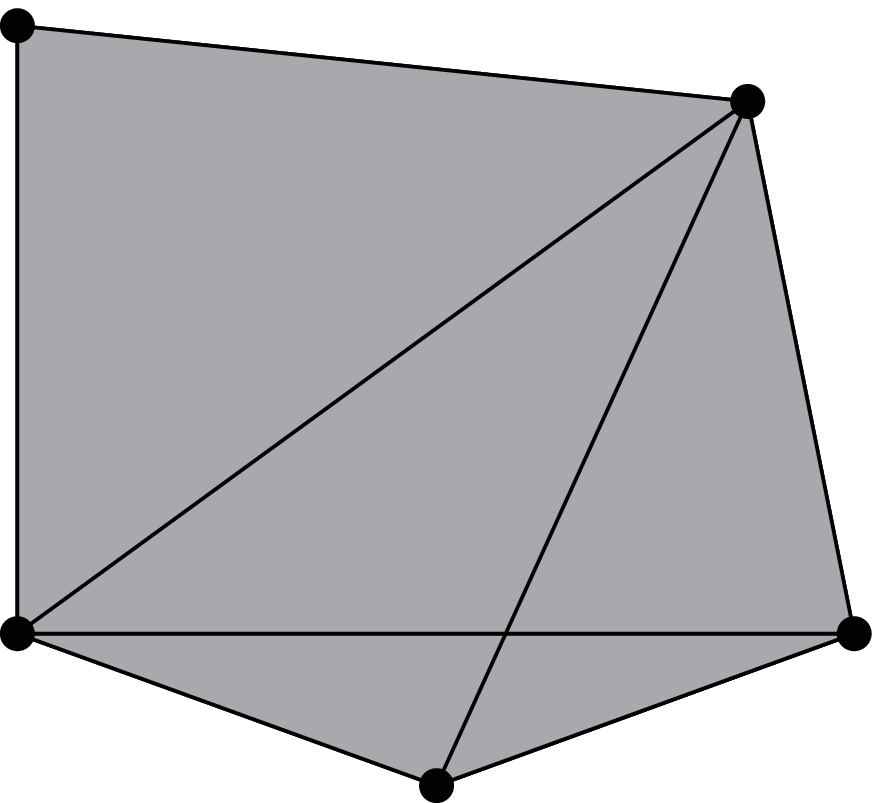}
\end{figure}

As an example, consider the set, $D$, of five points in the plane as pictured in Figures~\ref{data5} and~\ref{VRexamples}.  Each point in $D$ is a 0-simplex, each line segment drawn between points is a 1-simplex, and each shaded triangle a 2-simplex. As the parameter $r$ increases beyond $r= 4$ the Vietoris-Rips complex will contain additional 2-simplices, a 3-simplex at $r=4.9$, and eventually a 4-simplex when $2r$ is equal to the diameter of $D$. Note that the abstract simplicial complex $VR(D, 4.9)$ in Figure~\ref{VRexamples} cannot be geometrically realized in $\mathbb{R}^2$ since it contains pairs of 2-simplices whose intersection is not a face of either simplex.

We note that Vietoris-Rips complexes for increasing radius values are always a nested family of simplicial complexes associated to $D$, that is the complexes satisfy $$VR(D, r_1) \subseteq VR(D, r_2) \textrm{ whenever } r_1 \leq r_2.$$ This nested feature of the complexes along with the functorial nature of homology are what give rise the the concept of persistence to be defined below.

Although the Vietoris-Rips complex is relatively straightforward to define and calculate, it can be computationally expensive when used with large point clouds.  There are economical alternatives to the Vietoris-Rips complex, such as the lazy witness complex introduced in~\cite{carlsson}.  Persistence homology can be applied using any nested family of complexes indexed by some parameter. 


\subsection{Homology}
The homology of a simplicial complex $K$ is an algebraic measurement of how the $n$-simplices are attached to the $(n-1)$-simplices within $K$. 
Below we define some technical machinery (chains, boundary maps, and cycles) used to define homology groups, followed by some example calculations.

The {\it $p$-chains} of a simplicial complex $K$, denoted $C_p(K)$, is the group of formal linear combinations of the $p$-simplices of $K$ with coefficients from $\mathbb{Z}_2$. (More general definitions of homology with ring coefficients can be found in the standard algebraic topology texts~\cite{edelsbrunner, hatcher}.) Since $\mathbb{Z}_2$ is a field, the $p$-chains of $K$ are a  $\mathbb{Z}_2$-vector spaces with basis the $p$-simplices of $K$.  For example, the chains of $VR(D,4)$ are the vector spaces $C_0(VR(D,4)) = (\mathbb{Z}_2)^5$,  $C_1(VR(D,4)) = (\mathbb{Z}_2)^6$, and $C_2(VR(D,4)) = \mathbb{Z}_2$. 


The \textit{boundary map}, denoted $\delta_p$, identifies each $p$-chain with its boundary, a $p-1$ chain. Each boundary map, $\delta_p:C_{p}\to C_{p-1}$, is a homomorphism and in the case of $\mathbb{Z}_2$ coefficients, as considered here, these maps are linear transformations. 

Notice that $\delta_p \circ \delta_{p+1}$ is the zero map as the boundary of a boundary is empty. This fundamental property of chain complexes ensures that the image of $\delta_{p+1}$ is a normal subgroup of the kernel of $\delta_{p}$. The collective sequence of boundary maps and chains, as shown below, is called a {\it chain complex}.

$$\dots \stackrel{\delta_n}{\to} C_n(K) \stackrel{\delta_{n-1}}{\to} \dots \stackrel{\delta_2}{\to} C_1(K) \stackrel{\delta_1}{\to} C_0(K) \stackrel{\delta_0}{\to} 0, $$

\begin{figure}[h]\caption{The complex $VR(D,4)$ with an ordering assigned to its 0, 1, and 2-simplices.}
\centering
\begin{tabular}{cc}
\includegraphics[width=0.4\textwidth]{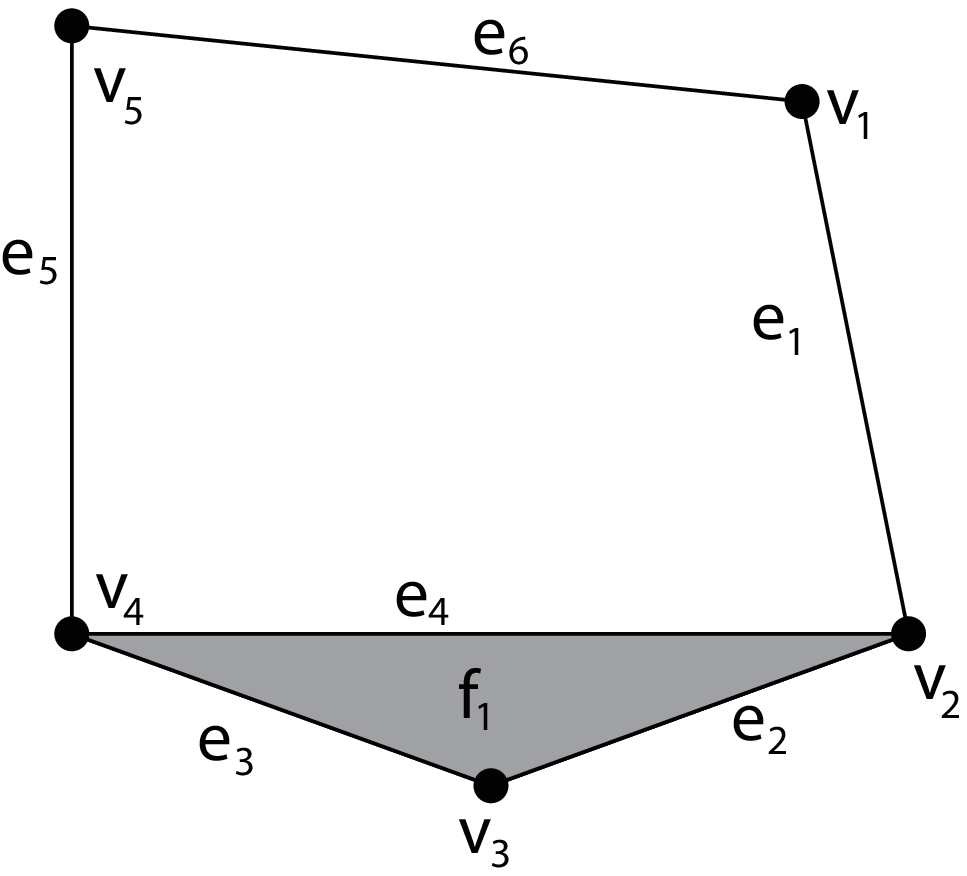}
\end{tabular}
\label{defnChains}
\end{figure}

Figure~\ref{defnChains} labels the simplices of $VR(D,4)$: the five 0-simplices, $v_1, v_2, v_3,$ $v_4$, $v_5$; six 1-simplices $e_1, e_2, e_3, e_4, e_5, e_6$; and one 2-simplex $f_1$.  With respect to this notation, the boundary of a chain is relatively easy to calculate. For example, $\delta_1 ( e_6+e_1+e_2 ) = v_5+v_3$ and $\delta_2 ( f_1) = e_2+e_3+e_4$.  More precisely, the chain complex of $VR(D,4)$ is

$$ 0 \longrightarrow \mathbb{Z}_2 \stackrel{\delta_2}{\longrightarrow} (\mathbb{Z}_2)^6 \stackrel{\delta_1}{\longrightarrow} (\mathbb{Z}_2)^5 \stackrel{\delta_0}{\longrightarrow} 0,$$

with boundary maps  given in matrix form by  $$\delta_2 = \left[ \begin{array}{c} 0 \\ 1\\ 1 \\ 1 \\ 0 \\ 0 \end{array} \right],  \delta_1=\left[
\begin{array}{cccccc}
 1 &  0 &  0& 0 &0  & 1  \\
 1 & 1  & 0 &1 &0  &0 \\
 0 &1   &1   &0 &0 &0 \\
 0 & 0 & 1 & 1 & 1 &0 \\
 0 & 0 & 0 & 0 & 1 & 1
\end{array}
\right], \textrm{ and } \delta_0 =  \left[\begin{array}{ccccc}0 & 0 & 0 & 0 & 0 \end{array}\right].
$$

Homology groups are defined using both the kernel and image of each boundary map. The kernel of $\delta_p$ is the set of all $p$-chains whose boundary is empty.  The elements of the kernel of $\delta_p$ are called {\it $p$-cycles} of $K$. The image of $\delta_{p+1}$ is the set of $p$-chains that are boundaries of a $p+1$-chain.  The {\it $p^{th}$ homology group of $K$}, denoted $H_p(K;\mathbb{Z}_2)$, is defined as the quotient group ker($\delta_p$)/image($\delta_{p+1}$).  

Intuitively, the $p^{th}$ homology group measures equivalence classes of  $p$-cycles of $K$ that are not ``filled" by $p+1$-chains. In homological dimension $p=1$ for the complex $VR(D,4)$, an example of a 1-cycle that is not the boundary of a 2-cycle is $e_1+e_2+e_3+e_5+e_6$. Hence this 1-cycle is in a non-zero equivalence class of $H_1(VR(D,4); \mathbb{Z}_2)$. The 1-cycle $e_2+e_3+e_4$, however, is the boundary of the 2-cycle $f_1$ (this 1-cycle is ``filled" by $f_1$), so this 1-cycle is equivalent to zero in the homology group.  Hence, in dimension $p=1$, the homology of $VR(D,4)$ is measuring the circular hole that is seen in the complex. 

To complete the homology calculation for the simplicial complex $VR(D,4)$ from Figure~\ref{defnChains}, we see
that the kernel of $\delta_0$ is $(\mathbb{Z}_2)^5$ and the rank of $\delta_{1}$ is four.  Thus $H_0(VR(D,4);\mathbb{Z}_2) \cong \mathbb{Z}_2$. Similarly, the nullity of $\delta_{1}$ is two and the image of $\delta_2$ is one dimensional. This implies that $H_1(VR(D,4); \mathbb{Z}_2) \cong \mathbb{Z}_2$. The group $H_2(VR(D,4);\mathbb{Z}_2)\cong 0$, since the kernel of $\delta_{2}$ is zero. Because the complex contains no simplices in higher dimensions, $H_p(VR(D,4);\mathbb{Z}_2)=0$ for all $p> 2$.

The calculation $H_0(VR(D,4); \mathbb{Z}_2) = \mathbb{Z}_2$ measures that $VR(D,4)$ is a connected complex.  The non-trivial group $H_1(VR(D,4);\mathbb{Z}_2)=\mathbb{Z}_2$ measures the existence of a one-dimensional cycle that is not the boundary of a 2-simplex, namely $e_1+e_2+e_3+e_5+e_6$. 

As the parameter $r>0$ increases the Vietoris-Rips complex includes more simplices,  thus the homology of the complex changes. For the complex $VR(D,4.9)$ the homology groups are $H_0(VR(D,4.9))=\mathbb{Z}_2$ and $H_p(VR(D,4.9))=0$ for all $p\geq 1$. In this example, the first homology group disappeared, or died, as $r$ increases from 4 to 4.9 as a result of the additional 2-simplicies that span the 1-cycle $e_1+e_2+e_3+e_5+e_6$.  

The functorial property of homology and the  inclusion map $i: VR(D, r_1) \rightarrow VR(D, r_2) \textrm{ whenever } r_1 \leq r_2,$ gives rise to induced maps between the homology of the complexes $i_*: H_*(VR(X,r_1);\mathbb{Z}_2) \rightarrow H_*(VR(X,r_2);\mathbb{Z}_2)$.  A nontrivial homology class $\alpha \in H_*(VR(X,r_1);\mathbb{Z}_2)$ is said to be born at radius $r_b$ if $r_b$ is the least radius value for which $H_*(VR(X,r_b);\mathbb{Z}_2)$ contains an element mapping onto $\alpha$ under the map $H_*(VR(X,r_b);\mathbb{Z}_2) \rightarrow H_*(VR(X,r_1);\mathbb{Z}_2)$. The homology class $\alpha$ is said to die at radius value $r_d$ provided that $r_d$ is the least radius value for which the class $\alpha$ maps to zero in the mapping  $H_*(VR(X,r_1);\mathbb{Z}_2) \rightarrow H_*(VR(X,r_d);\mathbb{Z}_2)$.  The topological feature that $\alpha$ represents is then said to have a birth and death ``time" corresponding to the radius values $r_b$ and $r_d$. We say that the class $\alpha$ persists over the interval $[r_b, r_d]$. Persistence homology of a data set $D$ is a cataloguing of the homological classes of the abstract simplicial complexes $VR(D,r)$ that persist for large intervals of radius values, $r$.

\begin{figure}[h]
\centering
\includegraphics[width=.53\textwidth]{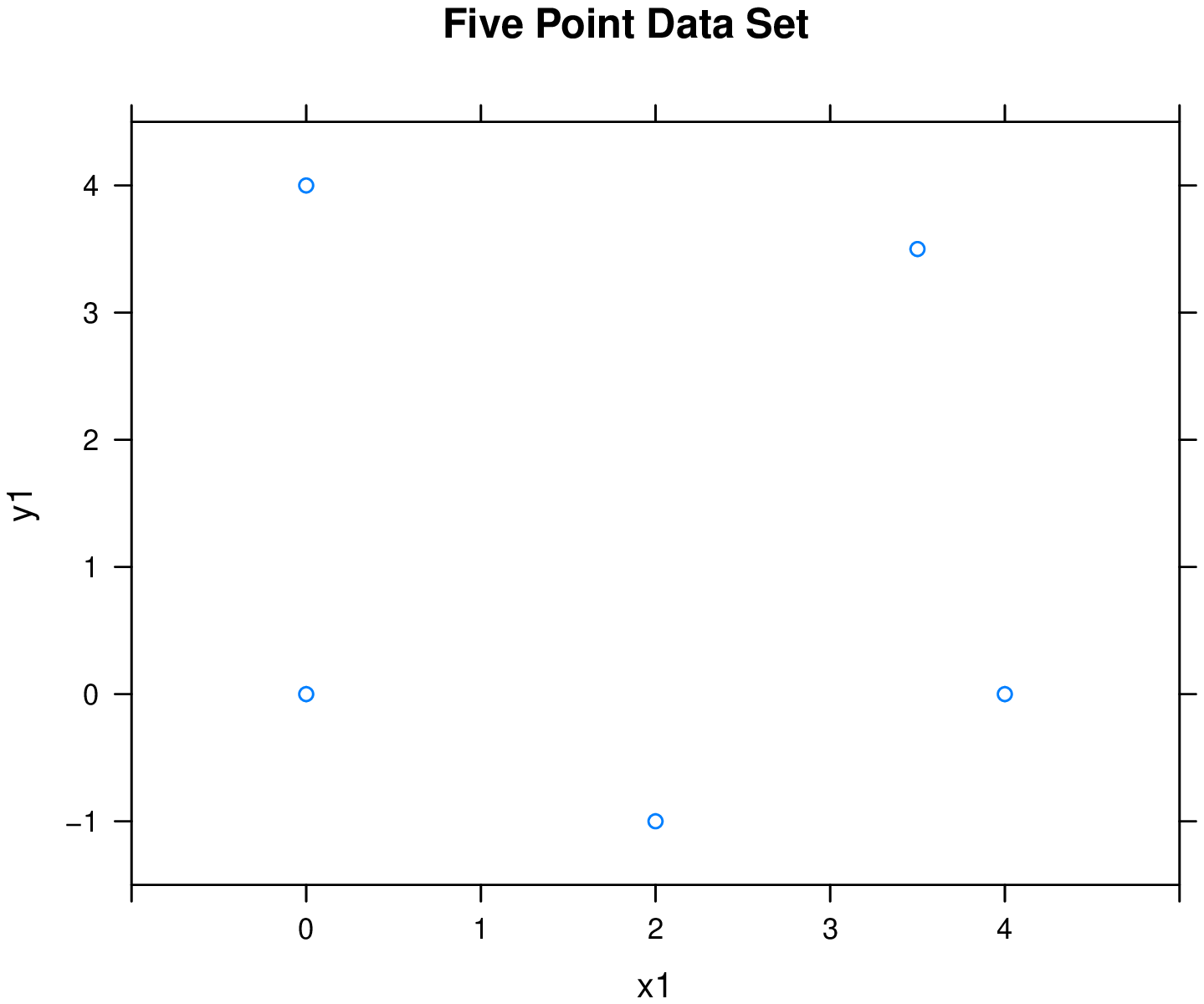} \hspace{.2in} \includegraphics[width=.48\textwidth]{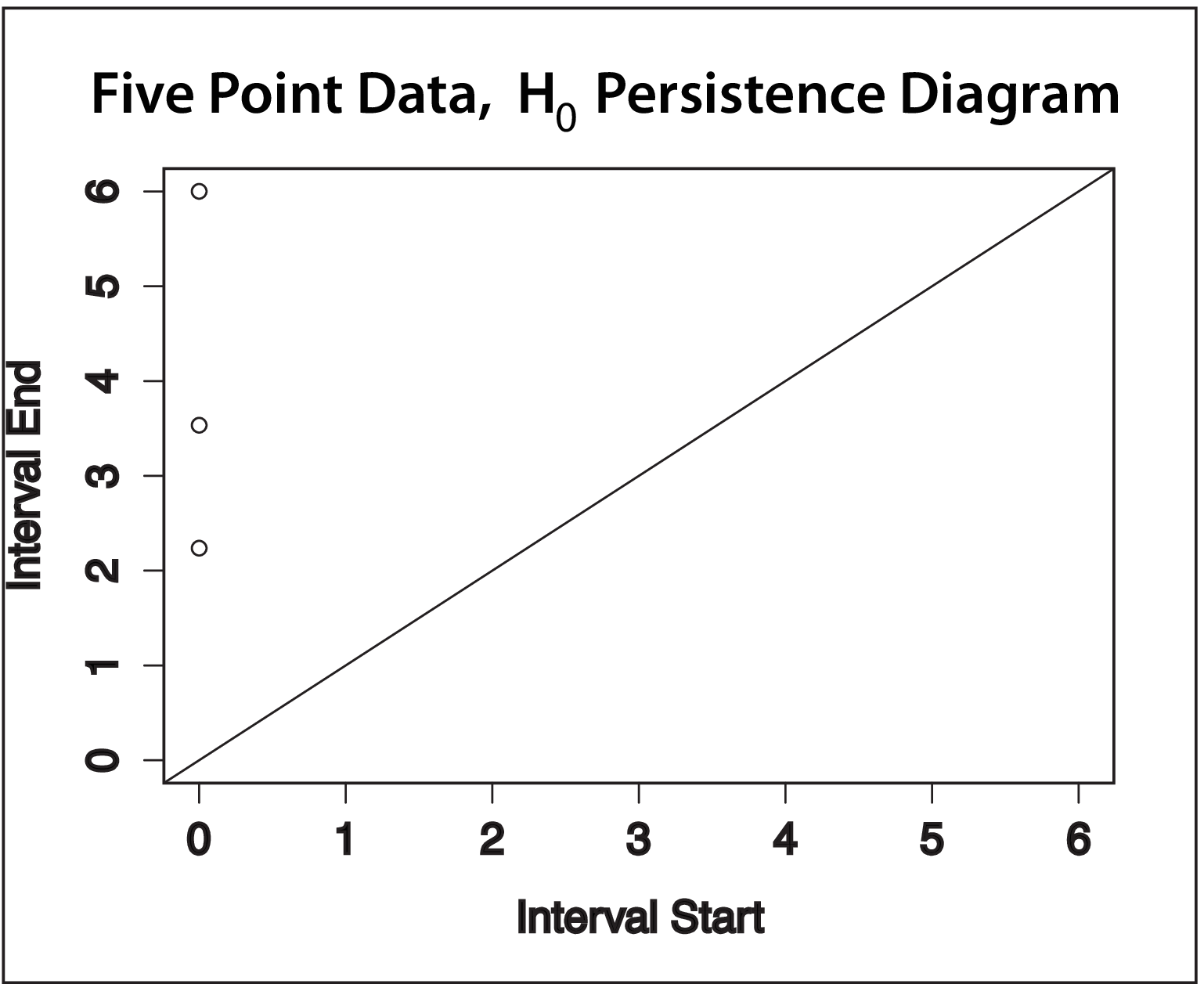}
\includegraphics[width=.48\textwidth]{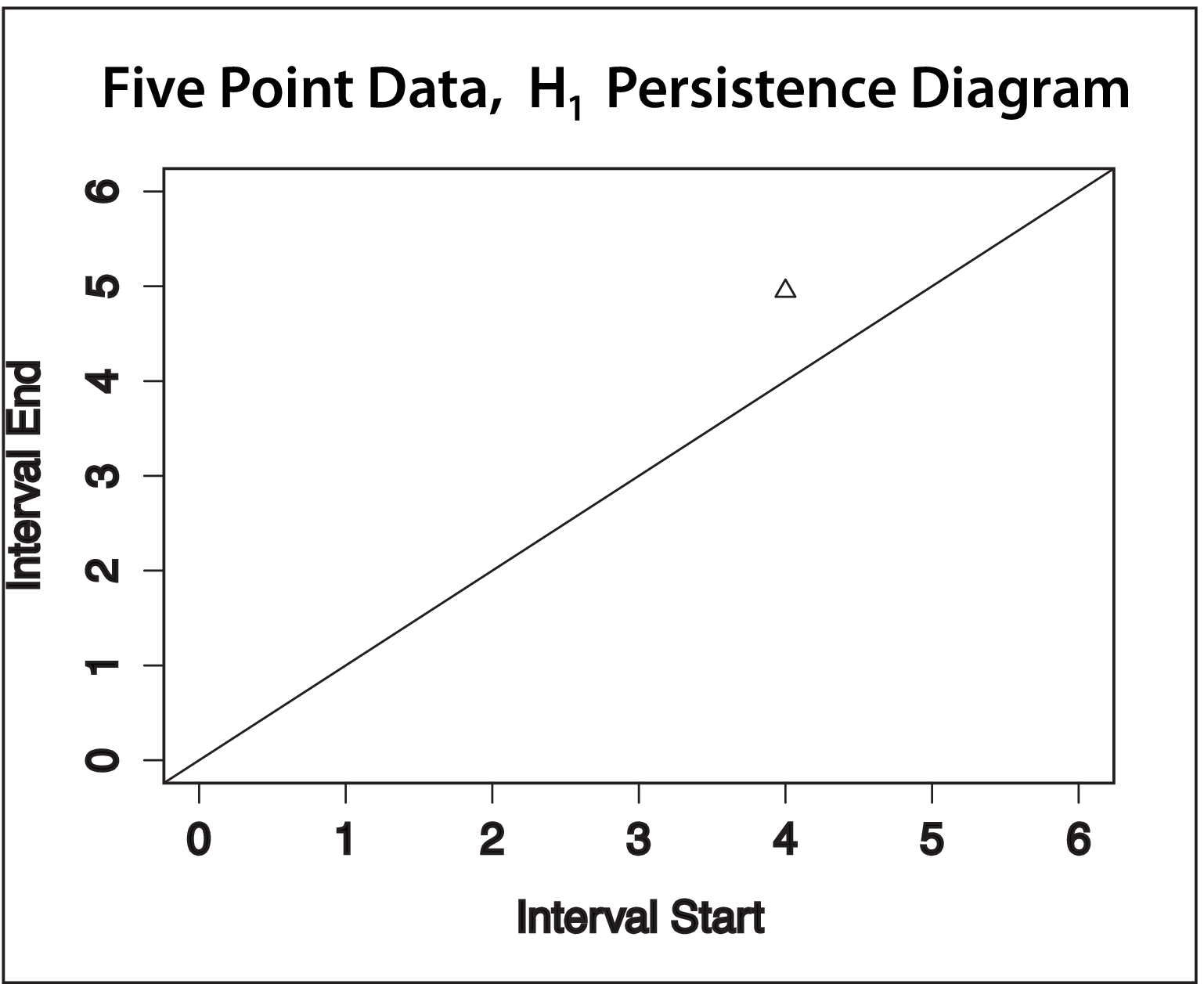}
\caption{A Five Point data set and the corresponding persistence homology diagrams in the homological dimensions 0 and 1.}
\label{pers-plot}
\end{figure}

For a fixed $k$, the {\it persistence diagram for $H_k(VR(X,*);\mathbb{Z}_2)$} is a plot of points $(r_b, r_d)$ for each non-zero class $\alpha \in H_k(VR(X,*); \mathbb{Z}_2)$. The persistence diagrams in Figure~\ref{pers-plot} display the $H_0$ and $H_1$ persistence diagrams for the five point data set $D$ first seen in Figure~\ref{data5}.  Note that all points in a persistence diagram are plotted above the line $y=x$ as a persistent homology class must be born before it can die.

Figure~\ref{pers-plot} includes a scaled version of the five point data set from Figure~\ref{defnChains} and the persistence diagram corresponding to this five point set. In homological dimension 1 (the $H_1$ diagram) the small triangle plotted at the point $(4,4.9)$ indicates that the five point data set contains a 1-dimensional homology class that is born at radius 4 and dies at radius 4.9.  In homological dimension 0 (the $H_0$ diagram) the circles plotted at the points (0, 2.236) and (0, 3.54) represent the connection of data points by 1-simplices at $r=2.236$ and at $r=3.54$ resulting in the ``death" of a connected component when it is joined with another connected component by a 1-simplex. For $r>3.54$ the five points are path connected via 1-simplices, thus this connected complex gives rise to a single 0-dimensional persistent homology class. This single class is plotted at $(0,6)$ as a result of considering only $r$-values in the range $0\leq r \leq 6$.  

\begin{figure}[h]
\centering
\includegraphics[width=.53\textwidth]{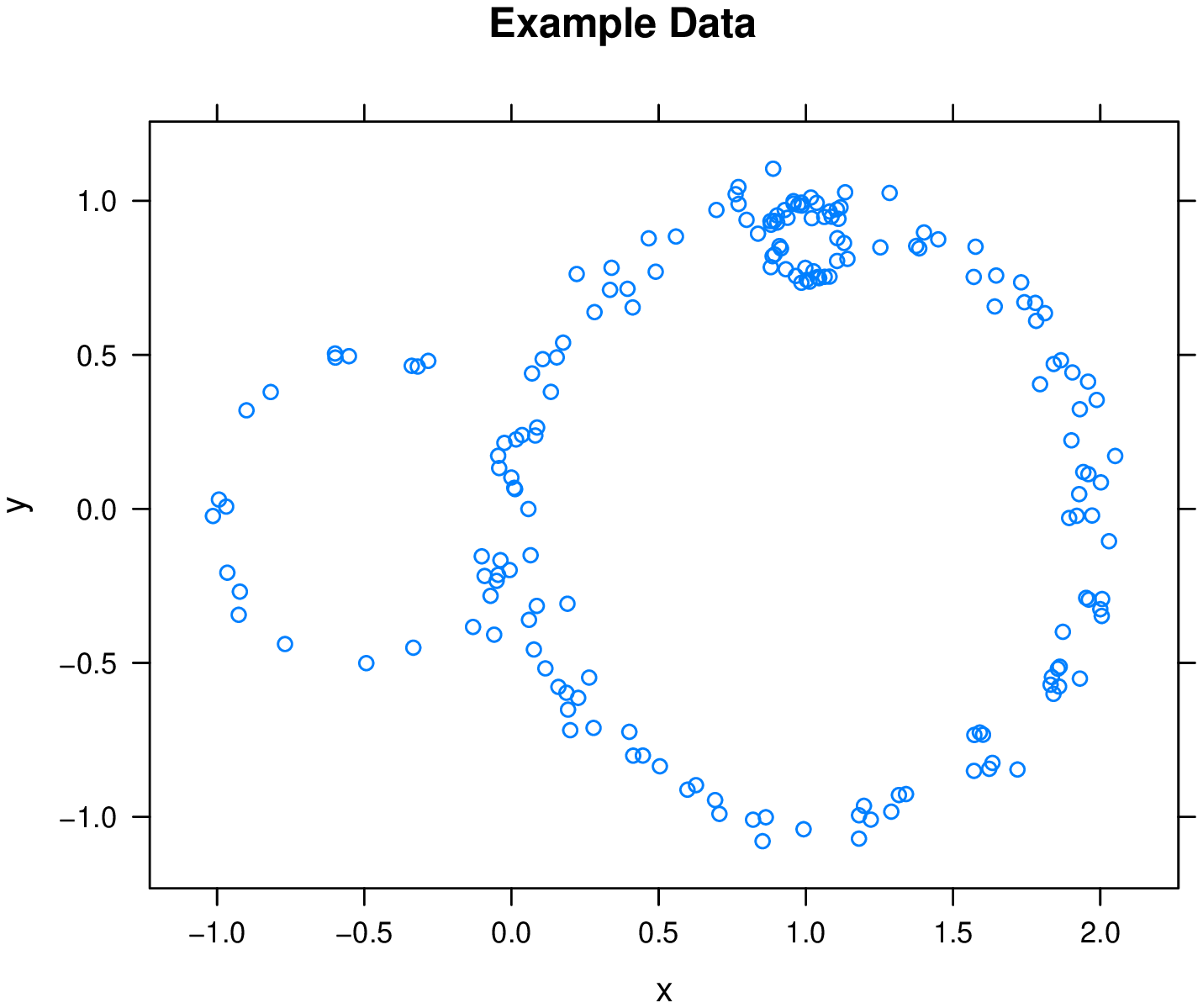} \hspace{.5in} \includegraphics[width=.48\textwidth]{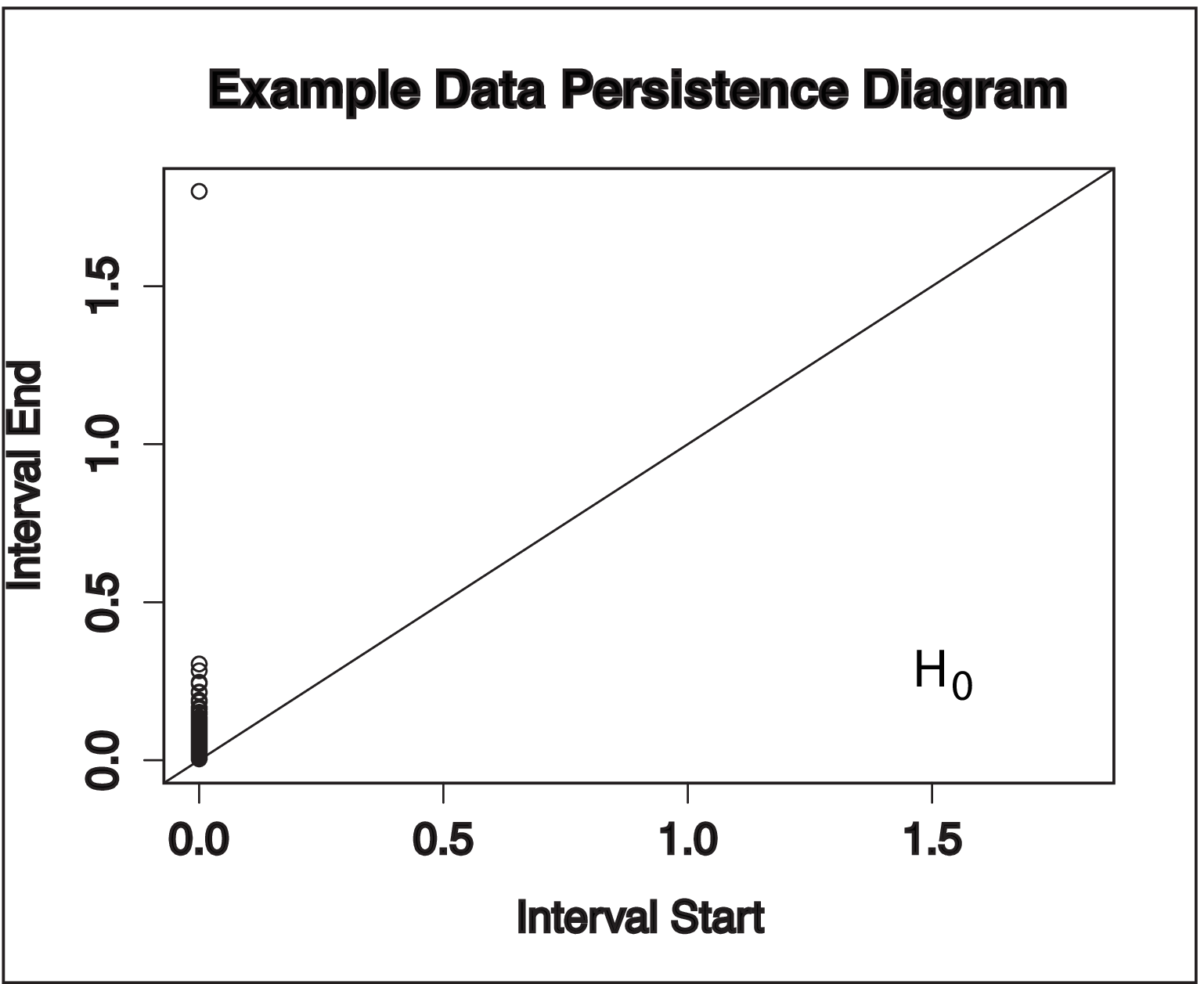}
\includegraphics[width=.48\textwidth]{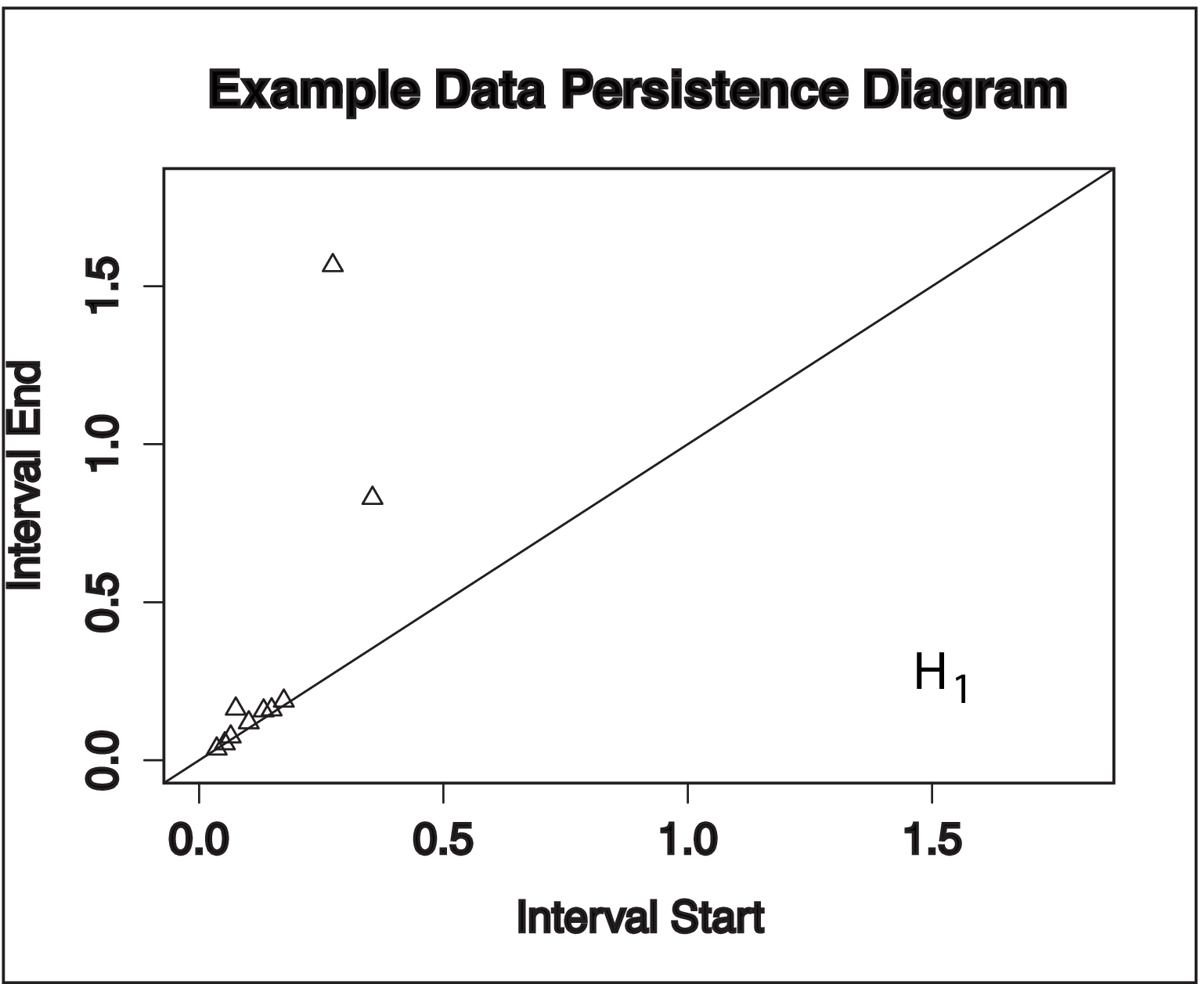}
\caption{An Example Data set and the corresponding  persistence homology diagrams for the homological dimensions 0 and 1.}
\label{pers-plot-morePts}
\end{figure}

Figure~\ref{pers-plot-morePts} contains a larger example data set that includes several 1-dimensional homological features of varying size and the persistence diagrams in dimensions zero and one corresponding to this example data set.

Within the persistence diagram in Figure~\ref{pers-plot-morePts}, we see two lone triangles at the points $p_1=(0.35, 0.8)$ and $p_2=(0.3,1.55)$.  The point $p_2$, with the early birth time, is the 1-dimensional homology class representing the larger circular feature on the right.  The earlier birth time is due to the closer scattering of the data points about the larger circle.  The point $p_1$, with the earlier death time, is the 1-dimensional homology class representing circle of smaller radius on the left. The early death time is due to the smaller radius of this circular feature.  The persistence diagram in Figure~\ref{pers-plot-morePts} also contains several triangles near the diagonal which represent classes that only peresist for a short while, and it includes a triangle at the point $(0.1,0.15)$ representing the 1-dimensional homology class resulting from the tiny circle of points at the top of the larger circle.  Notice that the 0-dimensional homology classes, which are plotted as small circles in the persistence diagram, all have birth time $r=0$ as a result of each data point representing a unique 0-dimensional class at $r=0$.  As $r$ increases, the complex consists of fewer connected components until it is one connected component.  The 0-dimensional persistence class plotted at the point $(0, 0.35)$ represents the joining of the last two components into a single component.  In other words, for $r\geq 0.35$ the simplicial complex $VR(X,r)$ is one connected component.  The 0-dimensional class plotted at $(0,2)$ is merely the result of using a maximum radius value of $r=2$ in the persistence homology calculation. This class indicates that the complex $VR(X,2)$ is one connected component.

\subsection{A Metric on Persistence Diagrams}
\label{PersistenceDiagramMetric}

We follow Robinson and Turner in selecting the metric on persistence diagrams that is analogous to the L2 norm in the space of functions on a discrete space.  Given two persistence diagrams $X$ and $Y$, let $x_1, x_2, \dots, x_n \in X$ be a listing of the off-diagonal points of $X$ and $y_1, y_2, \dots, y_m \in Y$ be the off-diagonal points of $Y$. Select points $x_{n+1}, \dots, x_{n+m}$ and $y_{m+1}, \dots, y_{m+n}$ along the diagonal so that $x_{n+k}$ is the point closest (in Euclidean distance) to $y_k$ and vise versa. Let $X'=\{x_1, \dots, x_{n+m}\}$ and $Y'= \{y_1, \dots, y_{n+m}\}$. We consider the set of all bijections $\phi: X' \rightarrow Y'$ such that (1) the off-diagonal point $x_k$ is paired either with an off diagonal point of $Y$ or with $y_{m+k}$ and (2) the diagonal point $x_l$ is paired either with $y_{l-n}$ or with one of the  diagonal points in $Y'$.  For a specific bijection $\phi$, if both $x_k$ and $y_j$ are diagonal points the {\it cost} of assigning $x_k$ to $y_j$, denoted $C(x_k, y_j)$, is 0, else the cost is the Euclidean distance between $x_k$  and $y_j$.

Define $d(X,Y)$, the {\it distance between the persistence diagrams $X$ and $Y$}, by
$$d(X,Y) = 
\left(
\inf_{\phi:X' \rightarrow Y'} \Sigma_{x\in X'} C( x,\phi(x)) \right)^{\frac{1}{2}}.$$

\begin{figure}[h]
\centering
\includegraphics[width=.4\textwidth]{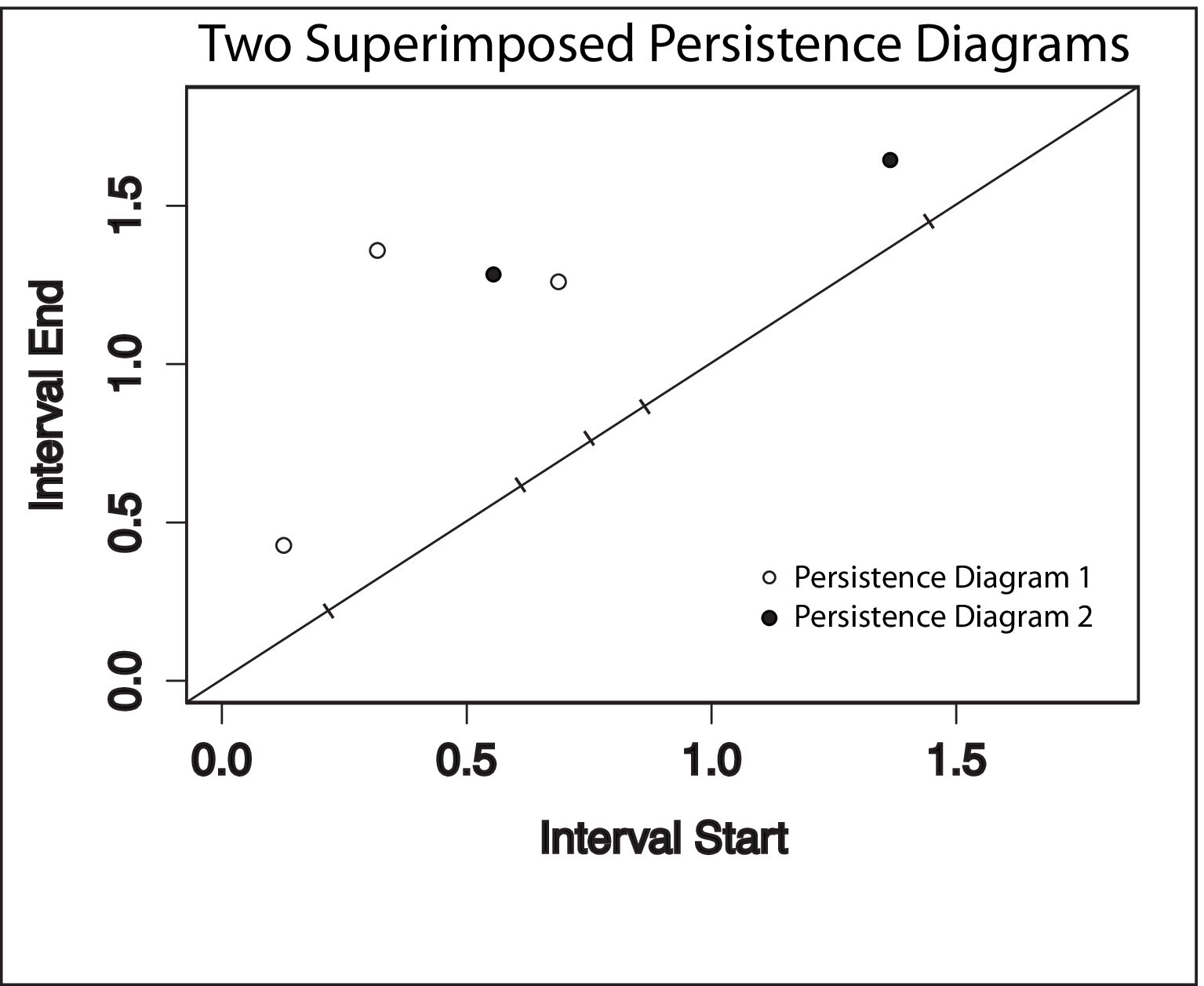} \hspace{0.1\textwidth}
\includegraphics[width=.4\textwidth]{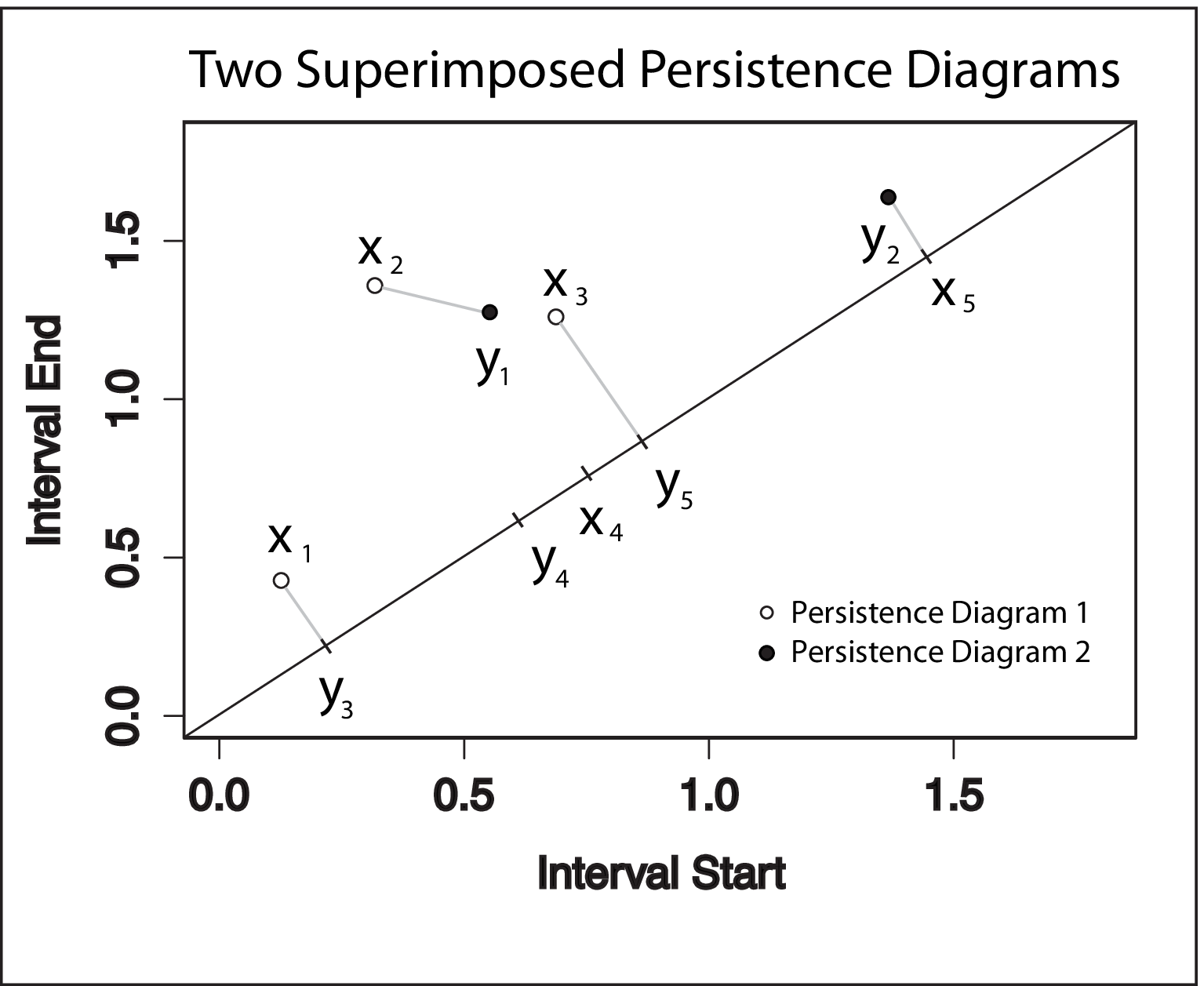}
\caption{On the left, two superimposed persistence diagrams of the same homological dimension.  On the right, the points $\{x_1,\dots, x_5\}$, $\{y_1, \dots, y_5\}$ and line segments indicating the optimal bijection. The diagram distance is the sum of the lengths of the line segments $\overline{x_1y_3}+\overline{x_2y_1}+\overline{x_3y_5}+\overline{x_5y_2}$.  The segment $\overline{x_4y_4}$ is not included as it is a segment between diagonal points. }
\label{diagram-distance}
\end{figure}

A bijection between $X$ and $Y$ is called optimal if it achieves the infimum.  The Hungarian Algorithm\cite{Hungarian, Munkres2}, also known as Munkres' assignment algorithm, presents a method for obtaining an optimal bijection in polynomial time. Figure~\ref{diagram-distance} gives an example of two simple persistent diagrams and the bijection exhibiting their diagram distance.

\section{Hypothesis Testing and Topological Data Analysis} 
\label{HTandTDA2spaces}

When persistence homology is applied to a random sample of points from a space, an element of variability is unavoidably introduced.  Different samples, if somewhat representative, are expected to have ``small" differences in their respective persistence diagrams, while samples from different spaces are expected to have comparatively ``large" differences in their persistence diagrams.  However, when the true shape-related features of two spaces are unknown, and all that is available are samples from each of these spaces, what qualifies as a ``small" or ``large" difference is unclear.  A tool is needed which can determine whether or not the shapes of such sampled spaces are measurably different.  Statistical hypothesis testing is a method that can be implemented in these situations to decide if there is sufficient evidence to classify the shapes of the spaces as measurably different.  A thorough development of statistical hypothesis testing is available in many standard sources~\cite{CasellaBurger, DegrootSchervish}.

\subsection{Hypothesis Testing via the Joint Loss Function}
Consider two spaces in $\mathbb{R}^m$, arbitrarily labeled $X_1$ and $X_2$, suspected of having measurably different shapes.  Suppose $n_1$ point-clouds are available from $X_1$ and $n_2$ point clouds are available from $X_2$, with their corresponding persistence diagrams in a fixed dimension denoted respectively by $X_{1,1},X_{1,2},$ $\ldots,$ $X_{1,n_1}$ and $X_{2,1},X_{2,2},\ldots,X_{2,n_2}$.  Further suppose that each of these $n_1+n_2$ point clouds was obtained via random sampling from either $X_1$ or $X_2$.  Within the statistical hypothesis testing paradigm, the null hypothesis asserts that the shapes of $X_1$ and $X_2$ are not measurably different, while the alternative hypothesis asserts the opposite.  The corresponding test statistic, proposed by Robinson and Turner \cite{RobinsonTurner}, is the \textit{joint loss function} $$\sigma^2_{\chi_{2}} = \sum_{m=1}^{2}\frac{1}{2n_m(n_m-1)}\sum_{i=1}^{n_m}\sum_{j=1}^{n_m}d(X_{m,i}, X_{m,j})^2,$$
where $d(\cdot , \cdot)$ is the persistence diagram distance metric described in Section \ref{PersistenceDiagramMetric}.

The joint loss function is ultimately an aggregate measure of within-group variation.  More specifically, $\sigma^2_{\chi_{2}}$ adds the variation in the ${n_1 \choose 2}$ persistence diagram distances from $X_1$ and the variation in the ${n_2 \choose 2}$ persistence diagram distances from $X_2$.  Unfortunately, the sampling distribution of $\sigma^2_{\chi_{2}}$ is non-trivial to determine and is currently unknown, which renders the ``standard" (i.e. distribution-based) hypothesis testing paradigm impossible.  To circumvent this, Robinson and Turner propose implementing a permutation test, which in this context is free of any distributional assumptions.  A thorough development of permutation tests is available in numerous sources~\cite{Higgins,RamseySchafer}.

To perform the permutation test, we assume that the null hypothesis is true, i.e. $X_1$ and $X_2$ are not measurably different in shape.  Such an assumption effectively means that the observed labeling of the sampled point clouds to either space $X_1$ or $X_2$ is just one of ${n_1+n_2 \choose n_1}$ possible assignments, all of which are arbitrary and equally likely.  For each of these possible assignments, the value of $\sigma^2_{\chi_{2}}$ is then computed.  Collectively, these values yield the \textit{permutation distribution} for $\sigma^2_{\chi_{2}}$, which is analogous to a sampling distribution in the standard hypothesis testing paradigm.  Finally, analogous to a standard hypothesis testing p-value, the \textit{permutation test p-value} is obtained by calculating the proportion of values in the permutation distribution which are less than or equal to the observed value of the joint loss function.  In practice the number of possible partitions may be considerably large, in which case a random sample of all possible partitions is selected and an \textit{approximate permutation test p-value} is obtained.

If the null hypothesis of the permutation test is actually false, then we would expect the permutation test p-value to be ``small" since the observed labeling of point clouds would be the only assignment that did not mix point clouds from both spaces.  When a permutation test p-value is less than some a-priori established threshold (e.g. $0.05$), the observed value of $\sigma^2_{\chi_{2}}$ is considered smaller than what can reasonably be explained by chance assignment of the sampled point clouds to spaces $X_1$ and $X_2$.  The null hypothesis would then be rejected and $X_1$ and $X_2$ classified as having measurably different shape.

It is important to note that if the sampled point clouds were not obtained via random sampling of $X_1$ and $X_2$, then a permutation test only allows us to draw conclusions with respect to the sampled point clouds.  For instance, if the permutation test p-value is less than our threshold, then we can conclude that the shapes of the sampled point clouds from $X_1$ and $X_2$ are measurably different; however, this conclusion cannot be generalized to $X_1$ and $X_2$ at large.  As limited as such a conclusion may be, it is still informative to know that such differences exist among the sampled point clouds, particularly when $m>3$ and the corresponding point clouds cannot be visualized.

\section{Extending Hypothesis Testing to Three or More Groups}
\label{HTandTDA3spaces}
While the methods of Section \ref{HTandTDA2spaces} are useful for determining whether or not two spaces are measurably different in a particular homological dimension, many practical applications involve more than two spaces.  The Cardiotocography data set considered in Section~\ref{CardioApp} is one such example.  Given $s\geq3$ spaces, suppose we have $n_1$ point clouds, obtained via random sampling, from space $X_1$, $n_2$ point clouds from space $X_2$, $\ldots$, and $n_s$ point clouds from space $X_s$.  In this section we extend the methods of Section \ref{HTandTDA2spaces} to obtain a hypothesis testing procedure which can determine whether or not sufficient evidence of measurable differences in shape exists between the $s$ spaces. 

\subsection{Hypotheses and Justification}
\label{HTandTDA3spacesTestingProcedure}
To conduct such an inquiry, we follow through with the suggestion of Robinson and Turner and use an approach analogous to a standard one-way ANOVA procedure in which there are potentially two stages of hypothesis testing.  An omnibus (i.e. ``global") test is conducted at the first stage and if this test produces significant results, a number of post-hoc (i.e. ``local") tests are performed at the second stage to identify the source(s) of the ``global" significance.  A thorough development of the one-way ANOVA procedure is available in several  sources~\cite{CasellaBurger, DegrootSchervish, RamseySchafer}.  As with the joint loss function in Section \ref{HTandTDA2spaces}, the sampling distribution of the test statistic corresponding to the omnibus test, which is presented below in Section \ref{HTandTDA3spacesOmnibusJLF}, is nontrivial to determine and currently unknown.  Hence, we again use a permutation test to carry out the omnibus test, which we will henceforth refer to as the \textit{omnibus permutation test}.  The logic behind and mechanics of this test are developed below in Section \ref{HTandTDA3spacesOmnibusJLF}.

The null hypothesis for the omnibus permutation test asserts that the shapes of $X_1$, $X_2, \ldots,$ $X_s$ are not measurably different, while the alternative hypothesis asserts that the shapes of at least two of the $s$ spaces are measurably different.  If we fail to reject the null hypothesis of this omnibus permutation test, then we are done.  However, if we reject the null hypothesis, then we know that at least two of the $s$ spaces have shapes that are measurably different, though we do not yet know which spaces.  Hence, up to ${s\choose2}$ post-hoc tests are performed, one for each possible pairing of two of the $s$ spaces.  For each post-hoc test, the null hypothesis asserts that the shapes of the two spaces are not measurably different, while the alternative hypothesis asserts that the shapes are measurably different.  Thus, each post-hoc test can be conducted via the methods described in Section \ref{HTandTDA2spaces}.

Before describing the test statistic and corresponding details for the omnibus permutation test, note that the purpose of the test pertains to type I error, which is the general term used to identify a hypothesis test decision in which the null hypothesis is incorrectly rejected.  To elaborate, an insignificant omnibus permutation test result prevents the analyst from unnecessarily performing any number of post-hoc tests.  Stated another way, if the null hypothesis of the omnibus permutation test is true, then all of the null hypotheses of the various post-hoc tests are also true, and thus do not need to be performed.  However, if an omnibus permutation test in which the null hypothesis is ultimately true is not performed, then $s-1$ post-hoc tests are unnecessarily performed and the chances of incorrectly rejecting the null hypothesis for at least one of these post-hoc tests, and then having to conduct additional unnecessary post-hoc tests, is dramatically increased.

\subsection{Omnibus Permutation Test Specifics}
\label{HTandTDA3spacesOmnibusJLF}
Suppose $n_1$ point-clouds are available from $X_1$, $n_2$ point clouds from $X_2, \ldots,$ and $n_s$ point clouds from $X_s$, with their corresponding persistence diagrams in a fixed dimension denoted respectively by $X_{1,1},X_{1,2},$ $\ldots,$ $X_{1,n_1}$, $X_{2,1},X_{2,2},\ldots,X_{2,n_2}$, and $X_{s,1},X_{s,2},\ldots,X_{s,n_s}$.  Analogous to the test statistic for the two-space permutation test presented in Section \ref{HTandTDA2spaces}, the test statistic for the omnibus permutation test, for three or more spaces, is a function of the diagram distances for all ${n_1 \choose 2}$ pairings of persistence diagrams from $X_1$, all ${n_2 \choose 2}$ pairings of persistence diagrams from $X_2$, $\ldots$, and all ${n_s \choose 2}$ pairings of persistence diagrams from $X_s$.  In particular, the \textit{omnibus joint loss function} is defined as $$\sigma^2_{\chi_{s}} = \sum_{m=1}^{s}\frac{1}{2n_m(n_m-1)}\sum_{i=1}^{n_m}\sum_{j=1}^{n_m}d(X_{m,i}, X_{m,j})^2,$$
where $d(\cdot , \cdot)$ is again the persistence diagram distance metric described in Section \ref{PersistenceDiagramMetric}.  Analogous to $\sigma^2_{\chi_{2}}$, $\sigma^2_{\chi_{s}}$ is ultimately an aggregate measure of variability since the omnibus joint loss function adds the within-group variation of persistence diagram distances from each of the $s$ spaces.  As previously mentioned, the sampling distribution of $\sigma^2_{\chi_{s}}$ is nontrivial to determine and currently unknown; hence, we turn to the omnibus permutation test. 

The logic behind and the mechanics of this omnibus permutation test are analogous to the two-space permutation test described in Section \ref{HTandTDA2spaces}.  We assume that the null hypothesis is true, which effectively means that the observed assignment of the sampled point clouds to the $s$ spaces is just one of $\sum_{i=1}^{s-1} {\sum_{j=i}^{s}n_j \choose n_i}$ possible assignments, all of which are arbitrary and equally likely.  For each of these possible assignments, the value of $\sigma^2_{\chi_{s}}$ is then computed.  Collectively, these values yield the permutation distribution for $\sigma^2_{\chi_{s}}$.  Finally, the permutation test p-value is then obtained by calculating the proportion of values in the permutation distribution which are less than or equal to the observed value of $\sigma^2_{\chi_{s}}$.  If the number of possible assignments is unreasonably large, a random sample of all possible assignments is selected and an approximate permutation test p-value is obtained.

Analogous to the two-space scenario of Section \ref{HTandTDA2spaces}, if the null hypothesis of this omnibus permutation test is actually false, then we would expect the permutation test p-value to be ``small" since the observed partitioning of point clouds would be the only partition that did not mix point clouds across the $s$ spaces.  The permutation test p-value is then compared to some a-priori threshold, such as $0.05$.  If the permutation test p-value is smaller than this threshold, then the observed value of $\sigma^2_{\chi_{s}}$ is considered smaller than what can reasonably be explained by chance assignment of the sampled point clouds to the $s$ spaces.  The null hypothesis would then be rejected and at least two of the $s$ spaces are declared as having measurably different shape.  To then identify the source(s) of this difference, i.e. to determine which spaces have measurably different shape, a requisite number of post-hoc tests are conducted via the two-space methods of Section \ref{HTandTDA2spaces}.

\section{Simulation Study}
\label{HTandTDA3spacesSimStudy}
To confirm the two-space permutation test introduced by Robinson and Turner \cite{RobinsonTurner} and to validate our proposed generalization for three or more spaces, we conducted a large-scale simulation study.  Throughout the study, shape was measured via one dimensional persistence homology.  Three different scenarios were considered and all three consisted of three spaces ($s=3$).  For each scenario, a trial consisted of obtaining 20 point clouds, via random sampling, from each of the three spaces and then calculating the approximate omnibus permutation test p-value.  All approximate omnibus permutation test p-values were based on 100,000 randomly selected assignments of the 60 collective point clouds to the three spaces.  In the third and final scenario, each of the three possible post-hoc tests were additionally performed using the two-space permutation test described in Section \ref{HTandTDA2spaces}.  The corresponding approximate two-space permutation test p-values were based on 100,000 randomly selected assignments of the 40 collective point clouds to the two respective spaces.  A total of 100 trials were performed for each scenario and the percentage of these 100 trials that produced approximate (omnibus/two-space) permutation test p-values less than or equal to 0.05 was calculated.

\subsection{Unbalanced Unit Circles}
\label{SimStudyUnbalancedUnitCircles}
For the first scenario, each of the three spaces was the unit circle; hence, the omnibus permutation test null hypothesis that there is no measurable difference in shape between the three spaces is ultimately true.  The sizes of the samples obtained from each space, however, were not equal (i.e. unbalanced).  Each point cloud in the ``first" space consisted of a random sample of size 18, whereas random samples of size 36 were obtained from the ``second" space and random samples of size 54 were obtained from the ``third" space.  For all three spaces, samples were obtained without allowing for measurement error; i.e. all sampled points lie on their respective unit circle.  Counter-intuitively, 100\% of the 100 trials performed produced approximate omnibus permutation test p-values less than or equal to 0.05.  In fact, 100\% of the trials produced approximate omnibus permutation test p-values less than 0.01.  Thus, in every trial the null hypothesis would be rejected at the 5\% level and we would conclude that the shapes of at least two of the three spaces are measurably different.

While such results may appear to suggest that the omnibus permutation test is ineffective, ultimately these results are an expected consequence of unbalanced sampling from the various spaces.  Relative to a random sample of size 18 from the unit circle, a random sample of size 54 is likely to produce a persistence diagram (corresponding to homology dimension one) containing a point that is measurably further from the diagonal. This point in the persistence diagram from a random sample of size 54 is expected as the circular feature within the sample will be "born" sooner and thus persist for a longer time interval. Hence, in order for the hypothesis testing methods described in Sections \ref{HTandTDA2spaces} and \ref{HTandTDA3spaces} to detect truly measurable differences in shape between the various spaces, balanced sampling must be implemented.

\subsection{Balanced Samples from Circles with Varying Radius}
\label{SimStudyBalancedCirclesVaryingFeatureSize}
For the second scenario, each of the three spaces was a circle with a radius of either 1,  1/2 or 1/3 units.  Notice that these three spaces are topologically equivalent, though geometrically different, and there is in fact a measurable difference in shape among the three spaces as measured by persistence homology in dimension one.  Hence, the null hypothesis for the corresponding omnibus permutation test is ultimately false.  Point clouds for each of the three circles consisted of random samples of size 24.  As in the Unbalanced Unit Circles scenario, all samples were obtained without allowing for measurement error; i.e. all sampled points lie on their respective circle.  Of the 100 trials performed, 100\% of them produced approximate omnibus permutation test p-values less than or equal to 0.05.  In fact, as in the Unbalanced Unit Circles scenario, 100\% of the trials produced approximate omnibus permutation test p-values less than 0.01.  Hence, in every trial the null hypothesis would be rejected at the 5\% level and we would conclude that the shapes of at least two of the three spaces are measurably different. 

As the three spaces of this second scenario are all topologically equivalent, these results suggest that the omnibus permutation test is capable of recognizing when purely geometrical differences exist between the spaces.  Stated another way, this second scenario suggests that the hypothesis testing methods described in Sections \ref{HTandTDA2spaces} and \ref{HTandTDA3spaces} are not scale invariant.  This is not a surprising result.  More specifically, as seen in the example data of Figure~\ref{pers-plot-morePts}, a sample from the circle with radius 1/3 will result in birth and death times for comparatively smaller radii values than a similar sized sample from the unit circle.  This is an artifact of the distances between neighboring points in the point cloud from the circle with radius 1/3 typically being smaller than those from the unit circle.   While in practice it will usually be difficult to determine whether a significant hypothesis test is a result of topological or geometrical differences between the various spaces, it is informative none the less to find evidence of any measurable difference in shape.

\subsection{Balanced Wedges}
\label{SimStudyBalancedWedges}
\begin{figure}[h]
\includegraphics[width=0.135\textwidth]{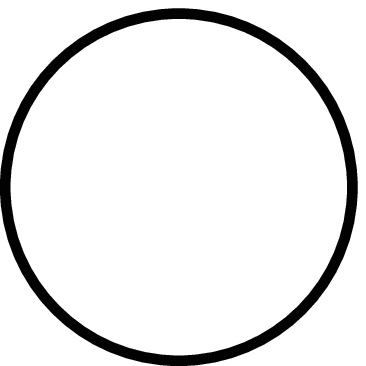} \hfill   
\includegraphics[width=0.27\textwidth]{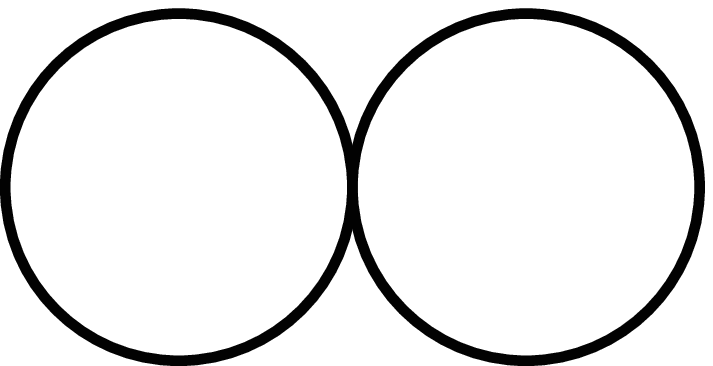} \hfill
\includegraphics[width=0.405\textwidth]{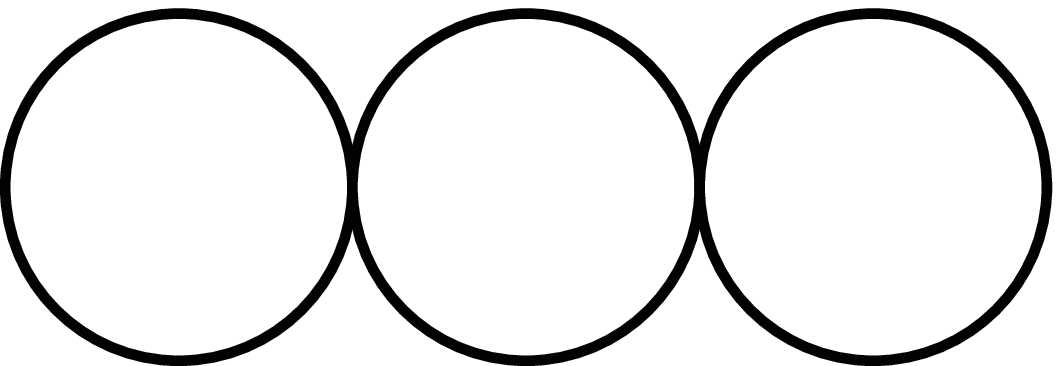}
\caption{Illustrations of the three spaces of case one within the Balanced Wedges simulation scenario.  On the left is the unit circle, in the middle is the wedge of two unit circles, and on the right is the wedge of three unit circles.}
\label{Figure:BalancedUnitWedgesSimulation}
\end{figure}

\begin{figure}[h]
\includegraphics[width=0.8\textwidth]{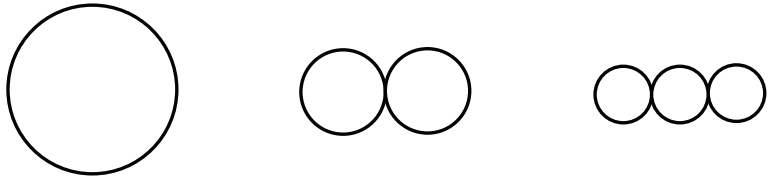}
\caption{Illustrations of the three spaces of case two of the Balanced Wedges simulation scenario.  On the left is the unit circle, in the middle is the wedge of two radius 1/2 circles,  and on the right is the wedge of three radius 1/3 circles.}
\label{Figure:BalancedScaledWedgesSimulation}
\end{figure}

\begin{figure}[h]
\includegraphics[width=0.8\textwidth]{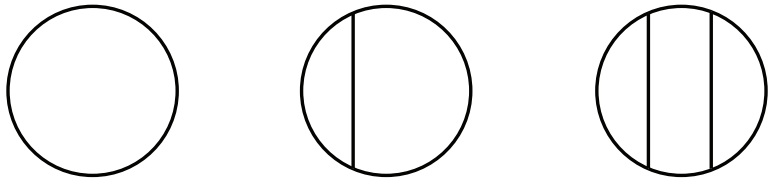}
\caption{Illustrations of the three spaces of case three of the Balanced Wedges simulation scenario.  On the left is the unit circle, in the middle is the unit circle with a single chord, and on the right is the unit circle with two non-intersecting chords.}
\label{Figure:BalancedCirclesChordsSimulation}
\end{figure}

The third and final scenario consisted of three distinct, but related cases in which only balanced sample sizes were considered. In the first case, the three spaces were the unit circle, the two-wedge consisting of two unit circles, and the three-wedge consisting of three unit circles.   Hence, in this first case, the radii of all component circles are one.  An image of these three spaces is given in Figure~\ref{Figure:BalancedUnitWedgesSimulation}.  In the second case, the three spaces are the unit circle consisting of one one-unit circle, the two-wedge consisting of two one-half unit circles, and the three wedge consisting of three one-third unit circles.  Hence, in this second case, the radii of the component circles within a space sum to one.  An image of these three spaces is given in Figure~\ref{Figure:BalancedScaledWedgesSimulation}. In the third and final case, the three spaces are the unit circle, the unit circle with a single chord traversing the interior of the circle, and the unit circle with two non-intersecting chords traversing the interior of the circle.  Hence, in this third case, the area of each of the three spaces is $\pi$ units.  An image of these three spaces is given in Figure~\ref{Figure:BalancedCirclesChordsSimulation}.  Observe that across these three scenarios the representations of the three spaces are topologically equivalent, but geometrically different.  We consider all three scenarios since persistence diagrams are unavoidably influenced by such differences.


\begin{figure}[h]
\includegraphics[width=0.25\textwidth]{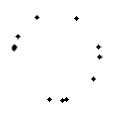}\hfill
\includegraphics[width=0.25\textwidth]{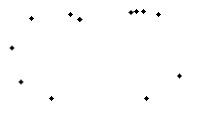}\hfill
\includegraphics[width=0.25\textwidth]{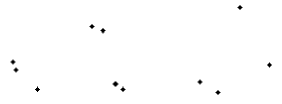}

\includegraphics[width=0.25\textwidth]{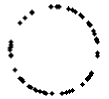}\hfill
\includegraphics[width=0.25\textwidth]{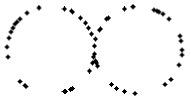}\hfill
\includegraphics[width=0.25\textwidth]{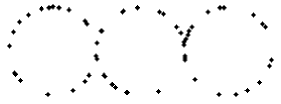}
\caption{Illustrations of random samples from each of the three spaces of case two of the Balanced Wedges simulation scenario.  The first row contains random samples of size 12.  The second row contains random samples of size 60.}
\label{Figure:BalancedUnitWedgesSamples}
\end{figure}

Within each of the three cases, the null hypothesis of the omnibus permutation test is ultimately false.  In other words, there are measurable differences in shape between the three spaces.  The point clouds for each of the three spaces, in all three cases, consisted of random samples of the same size (i.e. balanced samples).  Ten different sample sizes were considered: 6, 12, 18, 24, 30, 36, 42, 48, 54, and 60.  Figure~\ref{Figure:BalancedUnitWedgesSamples} provides examples of random samples of size 12 and 60, respectively, from each of the three spaces for case two.

\begin{figure}[h]
\begin{tabular}{ccc}

\includegraphics[height=3.0cm,width=4.0cm]{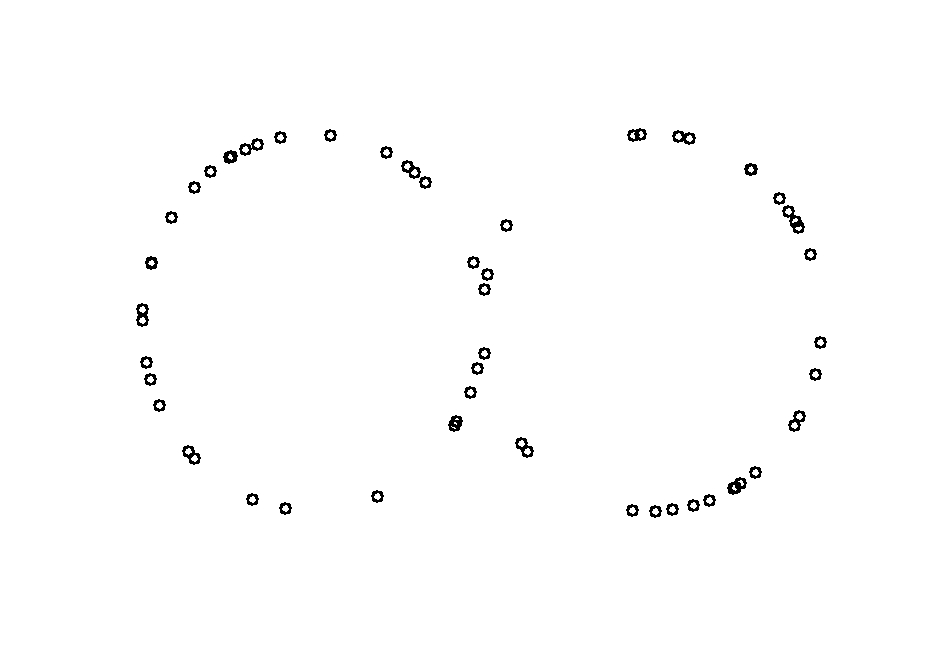}&
\includegraphics[height=4.0cm,width=4.0cm]{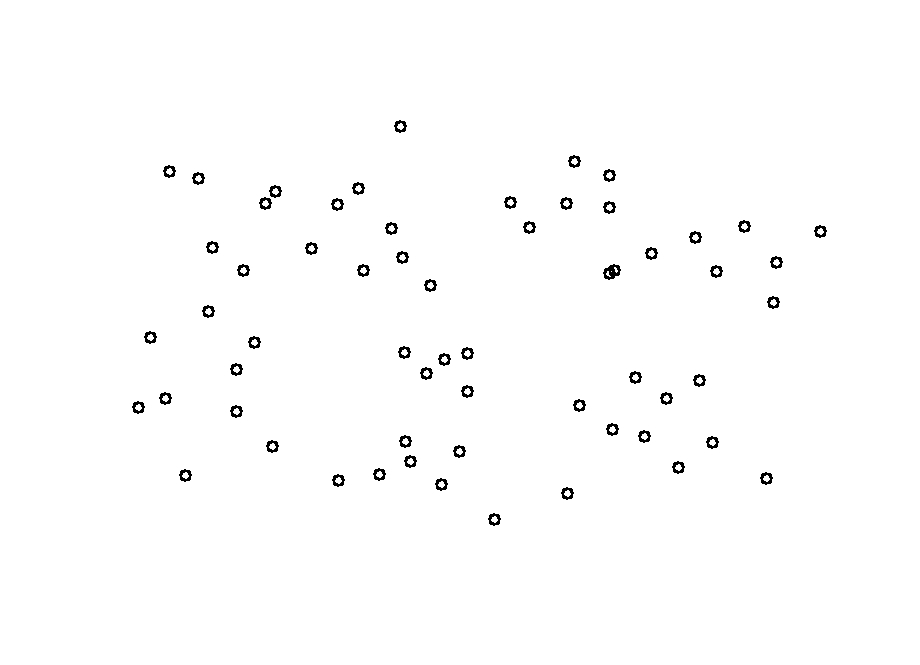}&
\includegraphics[height=4.0cm,width=4.0cm]{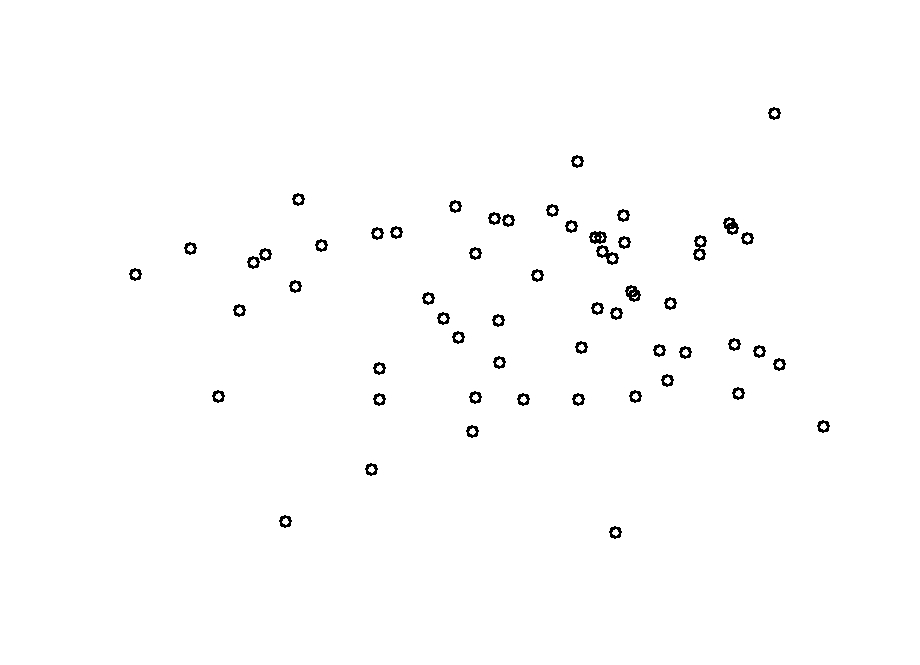}\\
\textrm{No Error} & \textrm{Error of 1/3} &  \textrm{Error of 2/3} 
\end{tabular}
\caption{Illustrative random samples of size 60, under various measurement errors, from the two-wedge of case one of the Balanced Wedges simulation scenario.  The measurement error, from left to right, is zero units, one-third units and two-third units.}
\label{Figure:TwoWedgeMeasurementError}
\end{figure}

For each of these ten sample sizes, three distinct measurement errors were considered: 0 (i.e. no error), 1/3, and 2/3 units.  For example, in the two-wedge of case one, measurement error was incorporated in the following manner.  A random sample of points was obtained separately from each of the two unit circles of the two-wedge.  Each point on either circle was obtained by randomly selecting the angle of the point from a Uniform(0,$2\pi$) distribution.  Each point was then assigned a radius value of 1 and converted to Cartesian coordinates.  Finally, for each point, two errors were randomly sampled from a Normal(0,$\sigma$) distribution, where $\sigma$ is the specified measurement error (e.g. 1/3), and respectively added to the Cartesian coordinates of the point.  For each of the three measurement errors, Figure \ref{Figure:TwoWedgeMeasurementError} exemplifies a sample of size 60 from the two-wedge.  From these images it is clear that as the measurement error increases, the extent to which the sample resembles the two-wedge dramatically decreases.  Measurement error for the other spaces of case 2, as well as for the other cases of scenario 3, were analogously incorporated. 

For each of the 30 combinations of sample size and measurement error, the percentage of the 100 trials producing an approximate omnibus permutation test p-value less than or equal to 0.05 for case one are given in Table \ref{table:BUW-OmnibusTests}.  Two trends are readily apparent from these results.  First, as sample size increases for a fixed measurement error, the percentage of significant omnibus permutation test results almost uniformly increases.  This is intuitive and desirable since we would expect measurable differences in shape between the three spaces to become more easily identifiable as sample size increases.  Second, as measurement error increases for a fixed sample size, the percentage of significant omnibus permutation test results almost uniformly decreases.  This too is intuitive and desirable since we would expect measurable differences in shape between the three spaces to become less easily identifiable as measurement error increases.  Given these trends and the fact that there are so many entries in the table at or near 100\%, these results suggest that the proposed omnibus permutation test ``successfully" identified  measurable differences in shape between at least two of these three spaces.  The results for cases two and three, as depicted in Figures \ref{Figure:BalancedScaledWedgesSimulation} and \ref{Figure:BalancedCirclesChordsSimulation}, are analogous to those above for case one and, therefore, are omitted.

\begin{table}[h]
\centering
\caption{Balanced Unit Wedges - Results of Omnibus Permutation Tests.  For each combination of sample size and measurement error, the percentage of approximate omnibus permutation test p-values (out of 100) yielding a value less than or equal to 0.05 is given.  The three populations are the unit circle, the two-wedge and the three-wedge.}
\begin{tabular}{|c|c|c|c|c|c|}
\hline
Sample Size & \multicolumn{3}{|c|}{Noise}\\\hline
 & 0 & $\frac{1}{3}$ & $\frac{2}{3}$  \\\hline
 6  & \au{6} & \au{9} & \au{1}  \\\hline
 12 & \au{95} & \au{57} & \au{18}  \\\hline
 18 & \au{100} & \au{65} & \au{41}  \\\hline
 24 & \au{100} & \au{96} & \au{41}  \\\hline
 30 & \au{100} & \au{100} & \au{85}  \\\hline
 36 & \au{100} & \au{100} & \au{98}  \\\hline
 42 & \au{100} & \au{100} & \au{100}  \\\hline
 48 & \au{100} & \au{100} & \au{100}  \\\hline
 54 & \au{100} & \au{100} & \au{100}  \\\hline
 60 & \au{100} & \au{100} & \au{100} \\
\hline
\end{tabular}
\label{table:BUW-OmnibusTests}
\end{table}

As the omnibus permutation test successfully identified measurable differences in shape between at least two of the three spaces, in all three cases, each of the three possible post-hoc tests were then conducted.  For each such post-hoc test, the null hypothesis asserts that there is no measurable difference in shape between the two spaces, while the alternative hypothesis asserts the opposite.  Hence, in all three tests, for all three cases, the null hypothesis is ultimately false.  As the results across the three cases were ultimately analogous, only the results for case one are discussed below.  In particular, for each of the 30 combinations of sample size and measurement error, the percentage of the 100 trials producing an approximate post-hoc test p-value less than or equal to 0.05 are given in Table \ref{table:BUW-1vs2} for the circle versus the two-wedge, in Table \ref{table:BUW-1vs3} for the circle versus the three-wedge, and in Table \ref{table:BUW-2vs3} for the two-wedge versus the three-wedge.

\begin{table}[h]
\centering
\caption{Balanced Wedges Case One - Results of Unit Circle vs. Two-Wedge Post-hoc Tests.  For each combination of sample size and measurement error, the percentage of approximate two-space permutation test p-values (out of 100) yielding a value less than or equal to 0.05 is given.}
\begin{tabular}{|c|c|c|c|c|c|}
\hline
Sample Size & \multicolumn{3}{|c|}{Noise}\\\hline
 & 0 & $\frac{1}{3}$ & $\frac{2}{3}$ \\\hline
 6  & \au{2} & \au{5} & \au{2} \\\hline
 12 & \au{90} & \au{29} & \au{13}  \\\hline
 18 & \au{99} & \au{40} & \au{15}  \\\hline
 24 & \au{100} & \au{83} & \au{28}  \\\hline
 30 & \au{100} & \au{97} & \au{49}   \\\hline
 36 & \au{100} & \au{100} & \au{64}  \\\hline
 42 & \au{100} & \au{100} & \au{80}  \\\hline
 48 & \au{100} & \au{100} & \au{82}  \\\hline
 54 & \au{100} & \au{100} & \au{92}  \\\hline
 60 & \au{100} & \au{100} & \au{97} \\
\hline
\end{tabular}
\label{table:BUW-1vs2}
\end{table}

\begin{table}[h]
\centering
\caption{Balanced Wedges Case One - Results of Unit Circle vs. Three-Wedge Post-hoc Tests.  For each combination of sample size and measurement error, the percentage of approximate two-space permutation test p-values (out of 100) yielding a value less than or equal to 0.05 is given.}
\begin{tabular}{|c|c|c|c|c|c|}
\hline
Sample Size & \multicolumn{3}{|c|}{Noise}\\\hline
 & 0 & $\frac{1}{3}$ & $\frac{2}{3}$  \\\hline
 6  & \au{2} & \au{5} & \au{1}  \\\hline
 12 & \au{97} & \au{65} & \au{30}  \\\hline
 18 & \au{100} & \au{85} & \au{40}  \\\hline
 24 & \au{100} & \au{100} & \au{53}  \\\hline
 30 & \au{100} & \au{100} & \au{95}  \\\hline
 36 & \au{100} & \au{100} & \au{100}  \\\hline
 42 & \au{100} & \au{100} & \au{100}  \\\hline
 48 & \au{100} & \au{100} & \au{100}  \\\hline
 54 & \au{100} & \au{100} & \au{100} \\\hline
 60 & \au{100} & \au{100} & \au{100}  \\
\hline
\end{tabular}
\label{table:BUW-1vs3}
\end{table}

\begin{table}[h]
\centering
\caption{Balanced Wedges Case One - Results of Two-Wedge vs. Three-Wedge Post-hoc Tests.  For each combination of sample size and measurement error, the percentage of approximate two-space permutation test p-values (out of 100) yielding a value less than or equal to 0.05 is given.}
\begin{tabular}{|c|c|c|c|c|c|}
\hline
Sample Size & \multicolumn{3}{|c|}{Noise}\\\hline
 & 0 & $\frac{1}{3}$ & $\frac{2}{3}$  \\\hline
 6  & \au{0} & \au{1} & \au{1}  \\\hline
 12 & \au{4} & \au{17} & \au{13}  \\\hline
 18 & \au{62} & \au{16} & \au{18}  \\\hline
 24 & \au{86} & \au{33} & \au{14}  \\\hline
 30 & \au{93} & \au{42} & \au{20} \\\hline
 36 & \au{87} & \au{66} & \au{26}  \\\hline
 42 & \au{95} & \au{67} & \au{43}  \\\hline
 48 & \au{99} & \au{87} & \au{65}  \\\hline
 54 & \au{100} & \au{93} & \au{66}  \\\hline
 60 & \au{100} & \au{98} & \au{84} \\
\hline
\end{tabular}
\label{table:BUW-2vs3}
\end{table}

The two trends that were apparent in the corresponding omnibus permutation tests for this simulation scenario are also readily apparent in all three of these post-hoc tests.  Specifically, as sample size increases for a fixed measurement error, the percentage of significant post-hoc tests tends to increase.  Similarly, as measurement error increases for a fixed sample size, the percentage of significant post-hoc tests tends to decrease.  A cell by cell comparison of the percentages among the three post-hoc tests, however, reveals an additional interesting trend.  The percentages for the post-hoc test between the circle and the three-wedge are almost uniformly larger than or equal to the corresponding percentages between the circle and the two-wedge, which are in turn almost uniformly larger than or equal to the corresponding percentages between the two-wedge and the three-wedge.  This too is mostly intuitive and desirable since, among the three spaces, the unit circle and the three wedge are the most different with respect to shape.  We are uncertain why the post-hoc test appears more adept at recognizing measurable differences in shape between the circle and the two-wedge rather than between the two-wedge and the three-wedge.  Regardless, all three of these trends, when coupled with the volume of entries in all three tables which are at or near 100\%, indicate that the proposed post-hoc tests ``successfully" identified measurable differences in shape between each of the three possible pairings of these three spaces.  Such findings additionally corroborate the legitimacy of the two-space permutation test.

\subsection{Summary of Findings}
\label{SimStudySummaryFindings}
In summary, the major findings of the simulation study are three-fold.  First and foremost, these simulations demonstrate that the proposed omnibus permutation testing procedure ``successfully" identified measurable differences in shape between at least two of the three spaces.  Second, these simulations confirm that the post-hoc testing component ``successfully" identified measurable differences in shape between any two spaces; such findings corroborate the legitimacy of the two-space permutation testing procedure.  Third and finally, these simulations reveal that these hypothesis testing procedures, for any number of spaces, require balanced sample sizes.

\section{Applications to Real Data Sets} 
\label{CardioApp}
%

We apply our methods to the Cardiotocography (CTG) data set that is freely available from the University of California at Irvine Machine Learning Repository.~\footnote{http://archive.ics.uci.edu/ml/datasets/Cardiotocography} 
The CTG data set includes 23 variables for each of 2126 subjects.  We apply our methods on a focused subset of four quantitative variables, including fetal heart rate baseline in beats per minute, number of accelerations per second, number of uterine contractions per second, and number of light decelerations per second.  These four quantitative variables are chosen because they are seemingly independent, and we want to consider no more than four such variables.  The categorical variable of interest is health status, which has three levels: normal, suspect, and pathologic.  The question of interest is whether or not the four-dimensional space created by the quantitative variables has a measurably different shape across the three health status groups.  To answer this question, we use the omnibus permutation testing procedure developed in Section 
~\ref{HTandTDA3spacesTestingProcedure}, measuring shape via one dimensional persistence homology.  Before this procedure can be performed, however, balanced samples from the three health status groups must be obtained.

Of the 2126 sampled subjects, 1655 are of normal health status, 294 of suspect health status, and 176 of pathologic health status. To obtain balanced samples across the three health status groups, we select a random sample of size 176 from both the normal and suspect health status groups. However, since the normal health status group is so large, we first test the representativeness of our sample of 176 subjects using an omnibus permutation test.  To that end, the 1655 normal health status subjects were randomly partitioned into nine ``spaces" of 176, leaving 71 discarded subjects.  The 176 subjects in each ``space" were then randomly partitioned into four four-dimensional point clouds of 44 subjects each.   The omnibus permutation test was then performed using these 36 point clouds.  The corresponding null hypothesis asserted that there were no measurable differences in shape between the nine ``spaces."  The resulting approximate permutation test p-value was based on 100,000 random assignments of the 36 point clouds to the nine ``spaces."  This entire process was then repeated 149 more times, where each such trial was based on a different initial random partition of the 1655 normal health status subjects into the nine ``spaces."  Of the 150 trials, 24 produced approximate permutation test p-values under 0.1, which suggests that in those trials there is evidence of measurable differences in shape between the nine ``spaces."  In other words, in those 24 trials, there is evidence that the nine ``spaces" may not be equally representative of the normal health status group, with respect to shape.  Hence, we select our random sample of 176 normal health status subjects by randomly selecting one of the nine ``spaces" from the ``more representative" 126 trials.

With balanced samples of 176 subjects from each of the three health status groups, we can now utilize the omnibus permutation testing procedure to address the original question of interest.   To that end, within each of the three health status groups, the 176 subjects were randomly partitioned into four point clouds of 44 subjects each.  The omnibus permutation test was then performed using the persistence diagrams corresponding to these 12 point clouds.  The corresponding null hypothesis asserted that there were no measurable differences in shape between the three spaces, i.e. health status groups.  The resulting permutation test p-value of approximately 0.003 was based on all 34650 possible assignments of the 12 persistence diagrams to the three spaces.  Given that the p-value is so small, we reject the null hypothesis and conclude that there are measurable differences in shape between at least two of the three spaces.

To determine the source(s) of the difference, we ultimately performed three post-hoc tests, one for each possible pairing of the three health status groups.  For each such test, the null hypothesis asserted that there were no measurable differences in shape between the two spaces of the respective health status groups.  All three resulting permutation test p-values were based on all 70 possible assignments of the 8 corresponding persistence diagrams to the two spaces.  For the normal and suspect health status groups, the permutation test p-value was approximately 0.029; for the normal and pathologic health status groups, the permutation test p-value was also approximately 0.029; for the suspect and pathologic health status groups, the permutation test p-value was approximately 0.257.  Hence, there is significant evidence of measurable differences in shape between the normal and suspect health status groups, and between the normal and pathologic health status groups, but insignificant evidence of such differences between the suspect and pathologic health status groups.

\section{Conclusion} 
\label{Conclusion}

For point clouds sampled from three or more spaces, we propose using an omnibus permutation test on the corresponding persistence diagrams to determine whether statistically significant evidence exists of measurable differences in shape between any of the respective spaces.  If such differences do exist, we then propose using a number of post-hoc (i.e. two-space) permutation tests to identify the specific pairwise differences.  To validate this proposed procedure, we conducted a large-scale simulation study using samples of point clouds from three distinct groups.  Various combinations of spaces, samples sizes and measurement errors were considered in the simulation study and for each combination the percentage of $p$-values below an alpha-level of 0.05 were provided.  The results of the simulation study clearly suggest that the procedure works, but additionally reveal that the method is neither scale invariant nor insensitive to imbalanced sample sizes across point clouds.  Finally, accounting for sample size and scale, we applied our omnibus testing procedure to a Cardiotocography data set and found statistically significant evidence of measurable differences in shape between the normal, suspect and pathologic health status groups. 

While the proposed ombinus testing procedure is applicable in any homological dimension, the simulation study and CTG application presented in this paper focus exclusively on homological dimension one.  Hence, to validate the effectiveness of the method in other homological dimensions, and to assess the consistency of the method across various dimensions, additional simulation studies can be performed.

\bibliographystyle{amsplain}

\end{document}